\documentclass{aa}
\usepackage{graphicx}
\usepackage{txfonts}
\newcommand{\Sc}[5]{\mbox{$#1\,^#2{\rm #3}^{{\rm #4}}_{\rm #5}$}}
\newcommand{\Se}{$S_{\!\!\rm e}$}
\newcommand{\SH}{$S_{\!\!\rm H}$}
\newcommand{\SP}{$S_{\!\!\rm P}$}
\begin{document}
\title{NLTE study of scandium in the Sun}
\author{H.W. Zhang
   \inst{1,2}
   \and
   T. Gehren\inst{2}
   \and
   G. Zhao\inst{3}
   }
\offprints{T. Gehren, \email{gehren@usm.lmu.de}}

\institute{Department of Astronomy, School of Physics, Peking University,
Beijing 100871, P.R. China \and Institut f\"ur Astronomie und Astrophysik der
Universit\"at M\"unchen, Scheinerstrasse 1, D-81679 M\"unchen, Germany \and
National Astronomical Observatories, Chinese Academy of Sciences, Beijing
100012, P.R. China\\}

\date{Received ; accepted }

\abstract
{}
{We investigate the formation of neutral and singly ionized scandium
lines in the solar photospheres. The research is aimed derive solar
$\log gf\varepsilon_{\odot}$(Sc) values for scandium lines, which
will later be used in differential abundance analyses of metal-poor
stars.}
{Extensive statistical equilibrium calculations were carried out for
a model atom, which comprises 92 terms for \ion{Sc}{i} and 79 for
\ion{Sc}{ii}. Photoionization cross-sections are assumed to be
hydrogenic. Synthetic line profiles calculated from the level
populations according to the NLTE departure coefficients were
compared with the observed solar spectral atlas. Hyperfine structure
(HFS) broadening is taken into account.}
{The statistical equilibrium of scandium is dominated by a strong
underpopulation of \ion{Sc}{i} caused by missing strong lines. It is
nearly unaffected by the variation in interaction parameters and
only marginally sensitive to the choice of the solar atmospheric
model. Abundance determinations using the ODF model lead to a solar
Sc abundance of between $\log\varepsilon_\odot = 3.07$ and $3.13$,
depending on the choice of $f$ values. The long known difference
between photospheric and meteoritic scandium abundances is confirmed
for the experimental $f$-values.}
{}

\keywords{line: formation - line: profiles - sun: abundances} \maketitle
%

\section{Introduction}

According to the theory of nucleosynthesis, the $\alpha$ elements are mostly
produced by Type II supernovae, while some iron-peak elements have significant
contributions from Type Ia supernovae. The synthesis process and sites of
scandium, as an element intermediate between $\alpha$ elements and iron-peak
elements in the periodic table, are not clear now.

The variation of the scandium abundance pattern in long-lived F-
and G-type stars with different metallicity can provide some
information on the element nucleosynthesis and the chemical
evolution of the Galaxy. There is an unresolved inconsistency
between different Sc abundance results. In some analyses, an Sc
\emph{enhancement} relative to Fe is found in metal-poor stars
(e.g. Zhao \& Magain \cite{zm90}, Nissen et al. \cite{ncsz00});
however, others (e.g. Gratton \& Sneden \cite{gs91}, Prochaska \&
McWilliam \cite{pm00}) found no evidence of any deviation from
[Sc/Fe] = 0.0.

Generally, the solar photospheric abundances serve as a reference
for abundance determinations in metal-poor stars, so a reliable
set of photospheric abundances is important. Ever since Anders \&
Grevesse (\cite{ag89}) published their widely used solar elemental
abundance table, many revisions and updates to photospheric and
meteoritic abundances of the elements have become available,
although the solar \emph{photospheric} scandium abundance has not
been updated for quite a long time. The photospheric abundance
value of log $\varepsilon$$_{\odot}\,({\rm Sc}) = 3.10 \pm 0.09$
adopted by Grevesse (\cite{g84}) was changed to $3.05 \pm 0.08$ by
Youssef \& Amer (\cite{ya89}). Neuforge (\cite{neu93}) obtained
$3.14 \pm 0.12$ from the \ion{Sc}{i} lines and $3.20 \pm 0.07$
from \ion{Sc}{ii} lines. The average value of $3.17 \pm 0.10$ was
adopted by Grevesse \& Noels (\cite{gn93}) and was kept in the
newest tabular version of Grevesse et al. (\cite{gas07}), which is
somewhat higher than the meteoritic value of $3.04 \pm 0.04$.

\begin{figure*}
\centering
\resizebox{15cm}{!}{\includegraphics{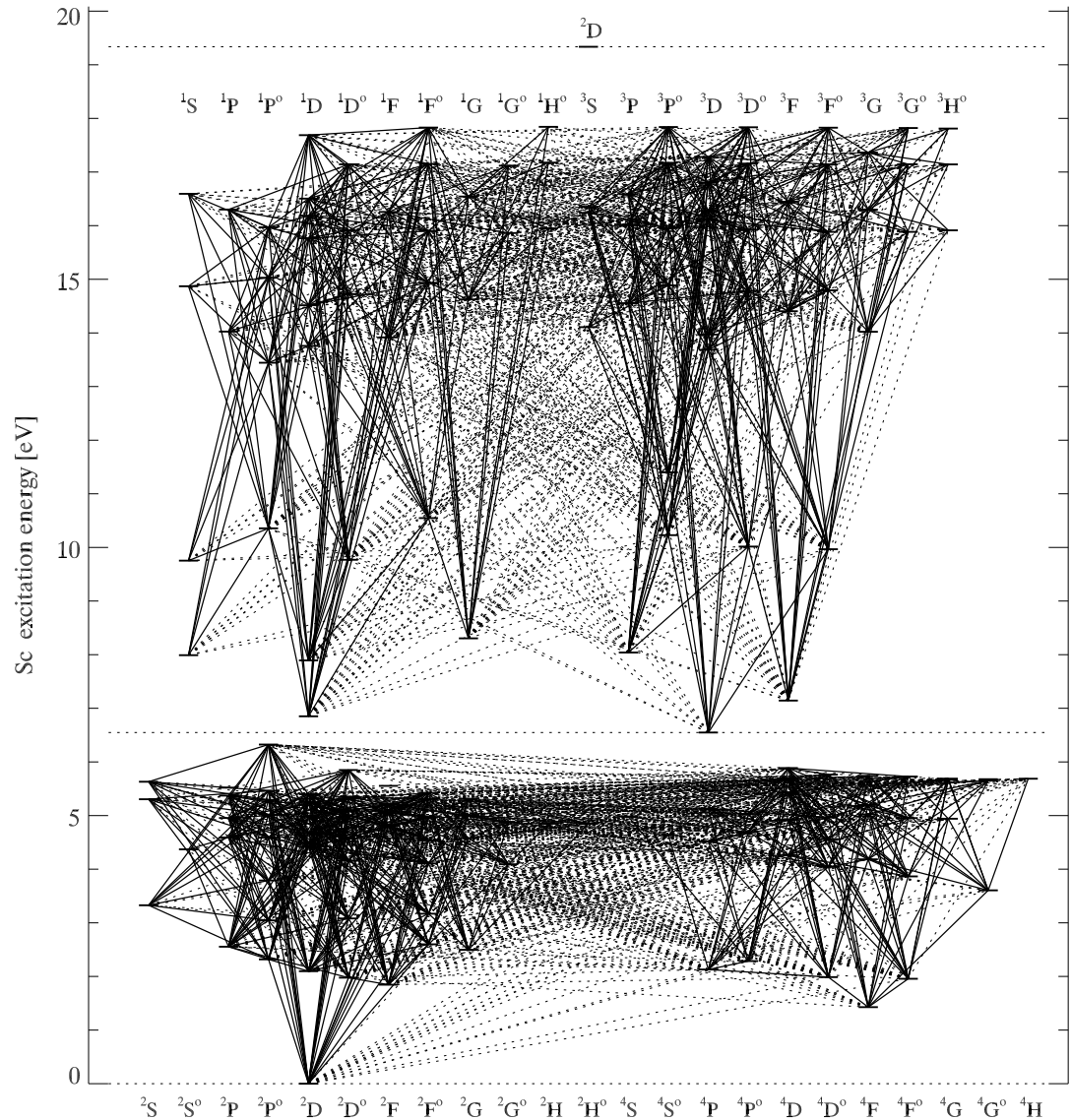}}\vspace{-7mm}
\caption[]{Adopted level structure of the \ion{Sc}{i} and
\ion{Sc}{ii} atom taken from the NIST data bank. Explicitly
calculated line transitions are indicated, where continuous lines
refer to allowed and dotted lines to forbidden transitions
according to selection rules assuming LS coupling.} \label{fig1}
\end{figure*}

It should be noted that local thermodynamic equilibrium (LTE) has
been assumed in previous papers about scandium abundance
determinations, and NLTE investigation of the scandium element has
never been published. In general, departures from LTE are
commonplace and often quite important, in particular for low surface
gravities or metallicities, with minority ions and low-excitation
transitions the most vulnerable (see the review paper of Asplund
\cite{as05}). In the Sun and other near-turnoff stars, the
ionization energy of \ion{Sc}{i} (6.56 eV, see Fig. \ref{fig1})
indicates that this is a \emph{minority ion}. Lines of \ion{Sc}{i}
should therefore be more susceptible to NLTE, because any small
change in the ionization rates largely changes the populations of
the minor ion, although there is no guarantee that \ion{Sc}{ii}
lines behave properly in a standard LTE analysis.

In this paper we investigate the statistical equilibrium and
formation of neutral and singly-ionized scandium lines in the solar
photosphere. We note that we do not intend to study the influence of
atmospheric inhomogeneities on any scale or that of chromospheric
temperatures and pressures. The method of NLTE calculations is
briefly introduced in Sect. 2. In Sect. 3 the synthesis of the
\ion{Sc}{i} and \ion{Sc}{ii} lines under NLTE and hyperfine
structure is presented. The discussion fills the last section.

\section{NLTE line formation calculations}

Abandoning the LTE approximation introduces a great deal of
additional complexity into the line formation calculations. Under
NLTE conditions, the atomic populations are described by a set of
statistical equilibrium equations in which radiative and collisional
processes are to be taken into account. Our calculations were
carried out with a revised version of the DETAIL program (Butler \&
Giddings \cite{BG85}), which solves the radiative transfer and
statistical equilibrium equations by the method of accelerated
lambda iteration.

Our calculations were performed on a partially ionized background
medium consisting of a plane-parallel, homogeneous, line-blanketed
theoretical model of the solar photosphere, which includes a simple
approach to convective equilibrium based on the mixing-length
theory. The model was computed with the MAFAGS code (Fuhrmann et al.
\cite{fu97}). In contrast to the line formation itself, the
\emph{atmospheric} model assumes LTE to obtain the final
temperature-pressure stratification. It uses opacity distribution
functions (ODF) for line-blanketing, based on Kurucz (\cite{ku92}),
and calculated with opacities rescaled to a solar iron abundance log
$\varepsilon_{\odot}$(Fe) = 7.51 (more details are found in Gehren
et al. \cite{GBMRS01}). The resulting atmospheric stratifications of
temperature and pressure are similar to those given by other solar
models (see comparison in Grupp \cite{gr04}, Fig. 15).

\subsection{Atomic model}
\label{amodel}

A comprehensive atomic model is required for NLTE calculations .
Similar to other iron-group elements, scandium has a complex atomic
structure. Our atomic reference model is constructed from 256 and
148 \emph{levels} for neutral and singly-ionized scandium,
respectively. Energies for these levels are taken from the NIST data
bank\footnote{http://www.physics.nist.gov/}. After a few early test
calculations with this complete fine structure model, we found that
the calculations could be considerably reduced by combining all fine
structure levels into 92 and 79 \emph{terms} for \ion{Sc}{i} and
\ion{Sc}{ii}, respectively. The corresponding fine structure data
were appropriately weighted to determine term energies. The atomic
term model used for our calculations is displayed in Fig.
\ref{fig1}. It shows that completeness is fading at high excitation
energies, with energy gaps of between 0.3 and 1.0 eV for the neutral
doublets and quartets, and gaps of $\sim$ 1.5 eV for the ionized
singlets and triplets.

The number of bound-bound \emph{transitions} treated in the NLTE
calculations is 1104 for \ion{Sc}{i} and 1034 for \ion{Sc}{ii},
numbers again considerably reduced from the original \emph{level}
transitions. Wavelengths and oscillator strengths of the fine
structure transitions are taken from Kurucz's database (Kurucz \&
Bell \cite{kb95}), and they are weighted by statistical weights to
yield artificial term transitions. A Grotrian diagram for
\ion{Sc}{i} and \ion{Sc}{ii} is displayed in Fig. \ref{fig1}. Solid
and dotted lines represent the allowed and forbidden transitions
included in the model atom, respectively. Since the reduction of the
fine structure model atom does not affect the resulting calculations
of the population densities, it is unlikely that the hyperfine
structure (HFS) has any direct influence on the NLTE results,
particularly since the known HFS is small.

For bound-free radiative transitions in the Sc atom, hydrogen-like
photoionization cross-sections (Mihalas \cite{mi78}) are adopted,
because data from the Opacity Project (OP; see Seaton et al.
\cite{se94}) are not available. In our current analysis this may be
the most uncertain representation. In previous studies of Fe (Gehren
et al. \cite{GBMRS01}), and K (Zhang et al. \cite{zg06}), where
complex calculations of such cross-sections were available, we found
that hydrogenic approximations occasionally tend to underestimate
the photoionization cross-sections by one or two orders of
magnitude. The effect on the NLTE analysis is examined below.

As usual, \emph{background opacities} are calculated with an opacity
sampling code based on the line lists made available by Kurucz
(\cite{ku92}). Since background opacities affect the photoionization
rates directly, their consideration is important. We note, however,
that the millions of faint lines, which may be somewhat more
important for model atmosphere construction, are marginal for our
line formation calculations.

In our calculations for Sc, we take into account inelastic
collisions with electrons and hydrogen atoms leading to both
excitation and ionization. Because laboratory measurements and
detailed quantum mechanical calculations for collision
cross-sections are absent, approximate formulae are applied. The
formulae of van Regemorter (\cite{re62}) and  Allen ({\cite{al73})
are used to describe the excitation of allowed and forbidden
transitions by electron collisions, respectively. Ionization
cross-sections for electron collisions are calculated with the
formula of Seaton (\cite{se62}). Drawin's (\cite{dr68}, \cite{dr69})
formula as described by Steenbock \& Holweger (\cite{sh84}) is used
to calculate neutral hydrogen collisions. Recently it was indicated
both experimentally and theoretically that Drawin's formula
overestimates the H collisional cross-section by one to six orders
of magnitude (e.g. Belyaev et al. \cite{be99}, Belyaev \& Barklem
\cite{bb03}), so a scaling factor $S_{\!\rm H}$ is applied to the
Drawin formula in our calculations, for which results are given
below.

\subsection{NLTE level populations}

\begin{figure}
\centering
\resizebox{\columnwidth}{!}{\includegraphics{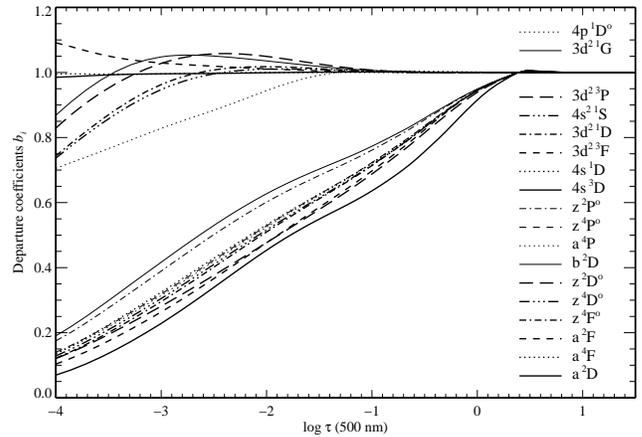}}\vspace{-0cm}
\caption[]{Departure coefficients $b_n$  for some levels of
\ion{Sc}{i} and \ion{Sc}{ii} in the ODF solar atmospheric model
applying our standard atomic model with multiplication factors \Se
= 1, \SH = 0.1, and \SP = 1 for bound-bound electron collisions,
bound-bound hydrogen collisions, and bound-free photoionization
coefficients, respectively. The kinetic equilibrium calculations
are described in the text.}\label{fig2}
\end{figure}

The atomic model described above still has a number of free
parameters that represent our ignorance of the microscopic
interaction processes. Whereas the number of levels (terms) and
lines (transitions) comprises the basic structure of the two lower
scandium ions, it is the interactions that require some additional
fine tuning. As explained above, we introduce standard
multiplication factors for electron collisions, hydrogen collisions,
and photoionization cross-sections. These factors are defined with
respect to the standard formulae for these three types of processes,
which were mentioned in section \ref{amodel}. Using the
\emph{standard atomic model}, departure coefficients $b_k = n_k^{\rm
NLTE}/n_k^{\rm LTE}$ for \ion{Sc}{i} and \ion{Sc}{ii} terms in the
solar atmosphere are presented in Fig. \ref{fig2}. The standard
model refers to \Se = 1, \SH = 0.1, and \SP = 1.
\begin{figure*}
\resizebox{\textwidth}{!}{\includegraphics{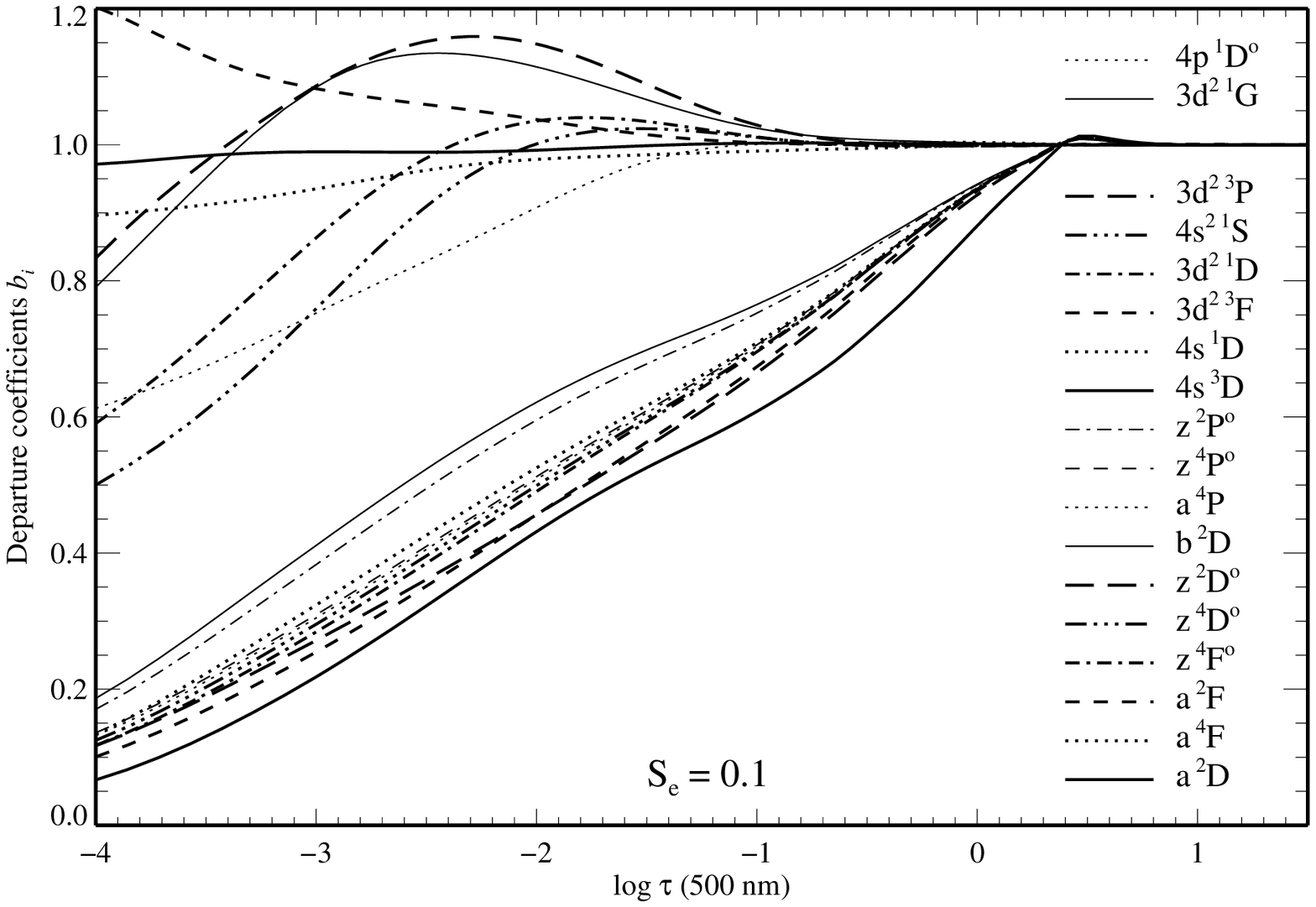}\includegraphics{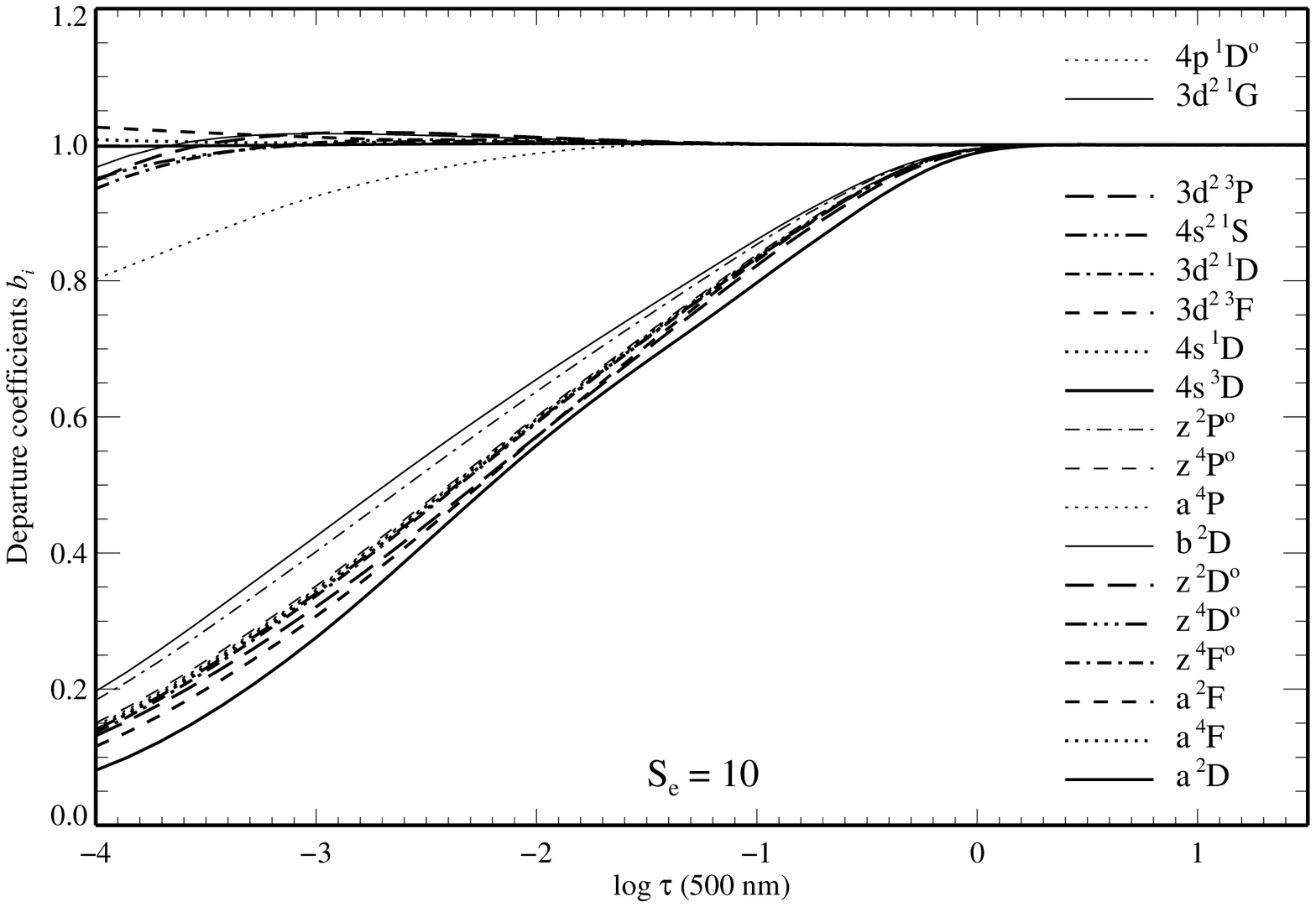}}\\
\vspace{-0.cm}
\resizebox{\textwidth}{!}{\includegraphics{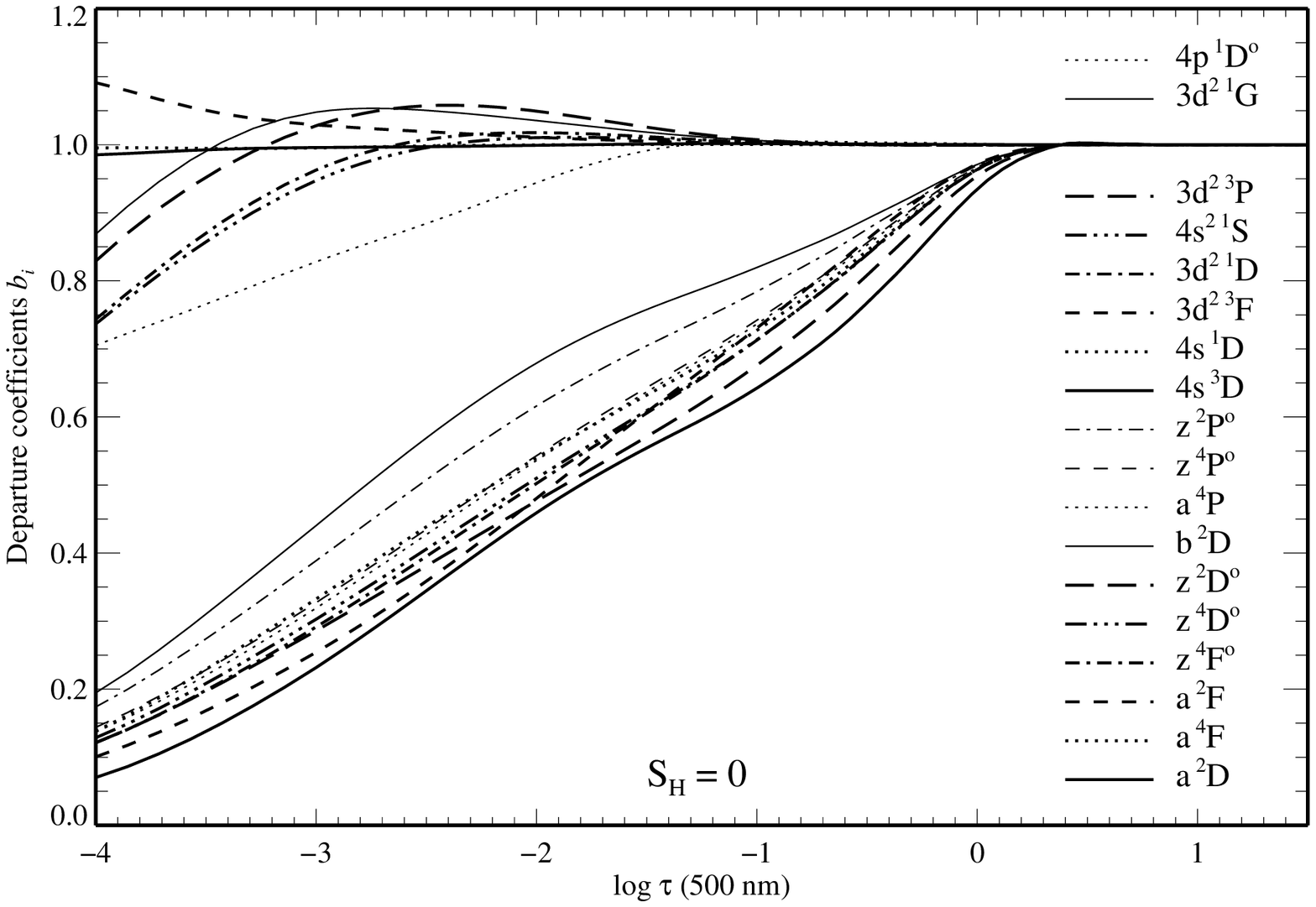}\includegraphics{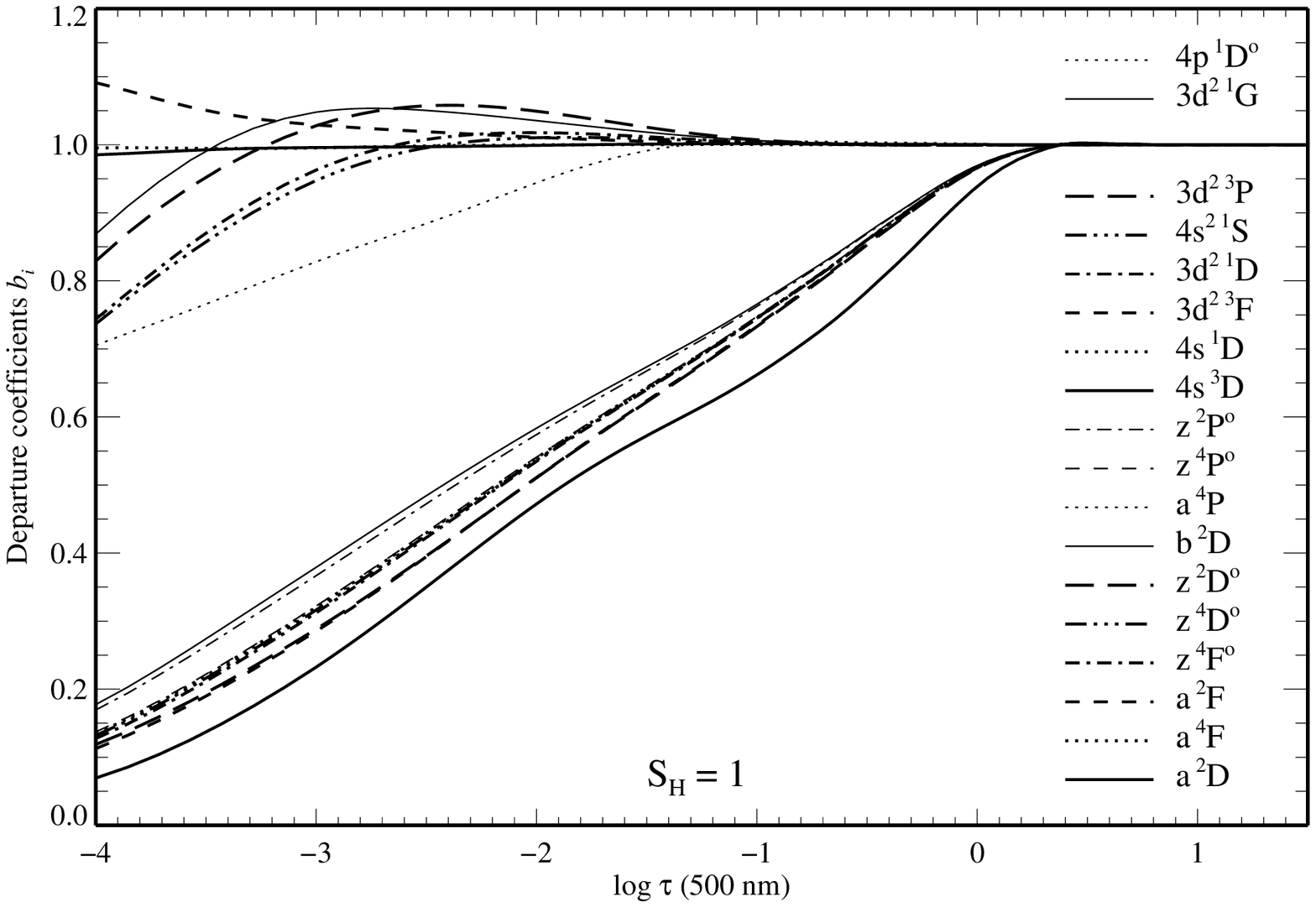}}\\
\vspace{-0.cm}
\resizebox{\textwidth}{!}{\includegraphics{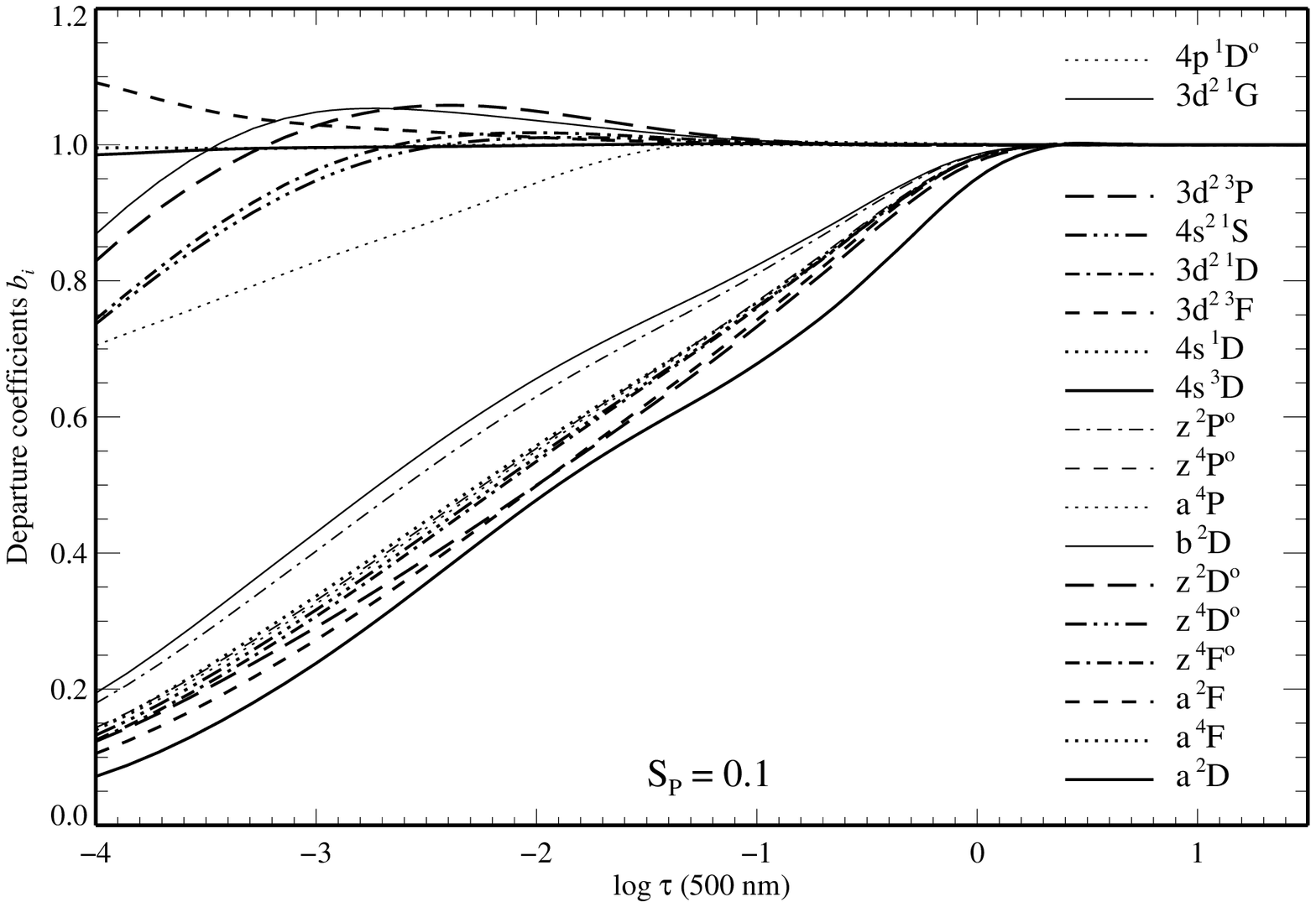}\includegraphics{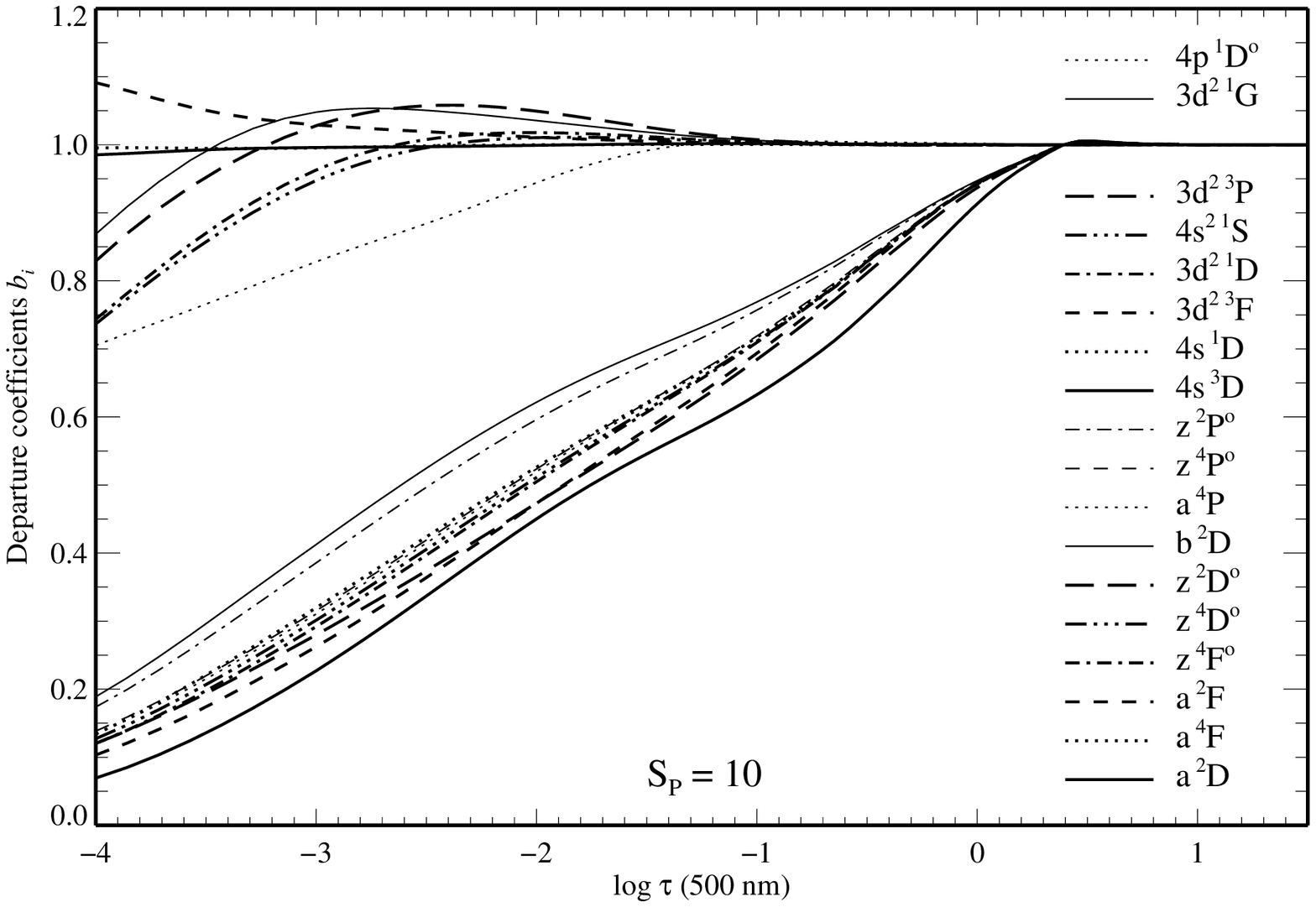}}
\vspace{-0.cm} \caption[]{Departure coefficients $b_n$  for levels
of \ion{Sc}{i} and \ion{Sc}{ii} in the ODF solar atmospheric model
testing the multiplication factors \Se = 0.1 and 10 (\emph{top}),
\SH = 0 and 1 (\emph{middle}), and \SP = 0.1 and 10
(\emph{bottom}).}\label{fig3}
\end{figure*}

\begin{table*}
\begin{minipage}{\linewidth}
\renewcommand{\footnoterule}{}
\caption{Atomic data and abundance results for \ion{Sc}{i} and
\ion{Sc}{ii} lines.}\label{table1}
\begin{center}
\tabcolsep0.15cm
\begin{tabular}{ccccrrrrrrrrrcrrc}
\noalign{\smallskip}\hline\noalign{\smallskip}
$\lambda$ & Mult  & Lower & Upper & $E_{\rm low}$ & $W_\lambda$ &  \multicolumn{2}{c}{log $gf$} & log $C_6$ & \multicolumn{2}{c}{$\log \varepsilon$} &
 $\Delta b$ & $\Delta X$ & log $gf \varepsilon_{\odot}$ & \multicolumn{3}{c}{HFS}\\
$[\AA]$   &        & level & level &  [eV]         &  $[m\AA]$   &Kurucz & LD~~ & &   ODF     &  OS    & &  & NLTE &  N  &  $\Delta\lambda$  &  -- \\
\noalign{\smallskip}\hline\noalign{\smallskip}
\ion{Sc}{i} & & & & & & & & & & & & & \smallskip\\
4023.690 &  5 & \Sc{a}{2}{D}{ }{5/2}  & \Sc{y}{2}{D}{o}{5/2}  & 0.02 &  51.9 &   0.405 &   0.381 & --31.780 &  3.04 &   3.05 &   0.00 &  +0.27 & 3.44 & 16 & 12 & *\\
5081.550 &  7 & \Sc{a}{4}{F}{ }{9/2}  & \Sc{y}{4}{F}{o}{9/2}  & 1.44 &  11.7 &   0.492 &   0.469 & --31.873 &  3.12 &   3.15 & --0.04 &  +0.16 & 3.61 & 22 & 86 & *\\
5671.800 &  6 & \Sc{a}{4}{F}{ }{9/2}  & \Sc{z}{4}{G}{o}{11/2} & 1.44 &  14.4 &   0.534 &   0.495 & --31.955 &  3.12 &   3.15 & --0.04 &  +0.15 & 3.66 & 21 & 95 &  \\
5686.830 &  6 & \Sc{a}{4}{F}{ }{7/2}  & \Sc{z}{4}{G}{o}{9/2}  & 1.43 &   9.1 &   0.415 &   0.376 & --31.955 &  3.03 &   3.06 & --0.07 &  +0.14 & 3.45 & 21 & 56 &  \\
\ion{Sc}{ii} & & & & & & & & & & & & & \smallskip\\
4246.830 &  1 & \Sc{4s}{1}{D}{ }{2}   & \Sc{4p}{1}{D}{o}{2}   & 0.32 & 160.6 &   0.283 &   0.241 & --32.088 &  3.05 &   3.09 &   0.00 & --0.04 & 3.33 & 13 & 21 & *\\
4314.087 &  5 & \Sc{3d^2}{3}{F}{ }{4} & \Sc{4p}{3}{D}{o}{3}   & 0.62 & 116.9 & --0.050 & --0.097 & --32.040 &  3.06 &   3.08 & --0.07 & --0.03 & 3.01 & 20 & 11 &  \\
4320.743 &  5 & \Sc{3d^2}{3}{F}{ }{3} & \Sc{4p}{3}{D}{o}{2}   & 0.60 & 108.2 & --0.212 & --0.252 & --32.040 &  3.05 &   3.08 & --0.03 & --0.06 & 2.84 & 15 &  9 &  \\
4400.350 &  4 & \Sc{3d^2}{3}{F}{ }{3} & \Sc{4p}{3}{F}{o}{3}   & 0.60 &  96.3 & --0.513 & --0.537 & --32.118 &  3.11 &   3.16 & --0.04 & --0.06 & 2.60 & 18 & 20 &  \\
4670.400 & 12 & \Sc{3d^2}{1}{D}{ }{2} & \Sc{4p}{1}{F}{o}{3}   & 1.35 &  63.5 & --0.518 & --0.576 & --31.996 &  3.03 &   3.04 & --0.04 & --0.04 & 2.51 & 15 & 17 &  \\
5031.020 & 11 & \Sc{3d^2}{1}{D}{ }{2} & \Sc{z}{1}{P}{o}{1}    & 1.35 &  75.4 & --0.341 & --0.400 & --32.081 &  3.06 &   3.06 & --0.03 & --0.03 & 2.72 &  9 & 20 & *\\
5357.200 & 19 & \Sc{3d^2}{3}{P}{ }{2} & \Sc{z}{1}{P}{o}{1}    & 1.50 &   4.3 & --2.143 & --2.111 & --32.068 &  3.16 &   3.17 & --0.01 &   0.00 & 1.02 &  9 &  4 & *\\
5526.810 & 21 & \Sc{3d^2}{1}{G}{ }{4} & \Sc{4p}{1}{F}{o}{3}   & 1.76 &  77.4 &   0.064 &   0.024 & --31.995 &  3.08 &   3.08 &   0.00 & --0.08 & 3.14 & 20 & 25 &  \\
5640.970 & 18 & \Sc{3d^2}{3}{P}{ }{1} & \Sc{z}{3}{P}{o}{2}    & 1.49 &  37.6 & --0.990 & --1.131 & --32.151 &  3.05 &   3.06 & --0.02 & --0.02 & 2.06 &  9 & 27 &  \\
5657.870 & 18 & \Sc{3d^2}{3}{P}{ }{2} & \Sc{z}{3}{P}{o}{2}    & 1.50 &  63.3 & --0.491 & --0.603 & --32.151 &  3.09 &   3.10 &   0.00 & --0.08 & 2.62 & 13 & 23 &  \\
5667.160 & 18 & \Sc{3d^2}{3}{P}{ }{1} & \Sc{z}{3}{P}{o}{1}    & 1.49 &  29.3 & --1.191 & --1.309 & --32.151 &  3.11 &   3.12 &   0.00 & --0.01 & 1.92 &  7 & 31 &  \\
5669.030 & 18 & \Sc{3d^2}{3}{P}{ }{1} & \Sc{z}{3}{P}{o}{0}    & 1.49 &  33.7 & --1.072 & --1.200 & --32.151 &  3.09 &   3.09 &   0.00 & --0.01 & 2.02 &  1 &  0 & *\\
5684.190 & 18 & \Sc{3d^2}{3}{P}{ }{2} & \Sc{z}{3}{P}{o}{1}    & 1.50 &  36.0 & --0.998 & --1.074 & --32.151 &  3.09 &   3.09 &   0.00 & --0.02 & 2.09 &  9 & 27 &  \\
6245.630 & 17 & \Sc{3d^2}{3}{P}{ }{2} & \Sc{4p}{3}{D}{o}{3}   & 1.50 &  35.7 & --1.030 &         & --32.057 &  3.02 &   3.03 &   0.00 & --0.02 & 1.99 & 15 & 37 &  \\
6279.760 & 17 & \Sc{3d^2}{3}{P}{ }{1} & \Sc{4p}{3}{D}{o}{2}   & 1.49 &  28.8 & --1.265 &         & --32.057 &  3.14 &   3.15 & --0.02 & --0.02 & 1.88 &  9 & 38 &  \\
6300.685 & 17 & \Sc{3d^2}{3}{P}{ }{2} & \Sc{4p}{3}{D}{o}{2}   & 1.51 &   7.3 & --1.887 &         & --32.057 &  2.98 &   3.00 & --0.04 & --0.01 & 1.09 & 13 & 33 &  \\
6320.867 & 17 & \Sc{3d^2}{3}{P}{ }{1} & \Sc{4p}{3}{D}{o}{1}   & 1.50 &   8.8 & --1.819 &         & --32.057 &  3.07 &   3.08 &   0.00 &   0.00 & 1.25 &  7 & 44 &  \\
\noalign{\smallskip}\hline\noalign{\smallskip}
\end{tabular}
\end{center}
\end{minipage}
\end{table*}

Since there are no other strong indicators, it is necessary to
explore the influence of such a parametric variation in interaction
strengths on the level population densities. This is done by varying
only one parameter at a time while holding all others at their
standard values, and the results are surprising in that we find no
strong influence for either of the multiplication factors (Fig.
\ref{fig3}). All calculations represented here document the extreme
weakness of the solar \ion{Sc}{i} lines resulting from both the low
abundance and the low ionization potential. All lines thus form in
the same atmospheric depths as the local H$^-$ continuum. There the
mean integrated line intensities become constant (and the \emph{net}
bound-bound radiative rates positive, or upwards) outside $\tau
\simeq 1$. All the \emph{lower levels} (those with less than $3
\ldots 4$ eV excitation energy) are therefore depopulated by this
collective pumping process, whereas the upper levels are
simultaneously populated. However, no true population inversion is
achieved. Photoionization rates from the lower levels are low, but
mostly higher than the corresponding recombination rates. This again
supports the depopulation of the lower levels throughout the solar
atmosphere. Of course, the radiative rates are modified by collision
rates, and the total rates are driven more towards zero net rates,
but this does not prevent the general depopulation trend.

Figure \ref{fig3} demonstrates clearly that even relatively large
variations in the multiplication factors do not change the run of
the level populations too much. There are thermalizing effects when
increasing electron or hydrogen collision factors, but these affect
the population \emph{ratios}, and not the departure coefficients
themselves very much. A notable exception is represented by the \Se
= 10, because this starts to couple the lower terms more efficiently
to the upper ones and thus thermalizes the whole \ion{Sc}{i} system.
The lack of variation with \SP\ is simply due to the decoupling of
the radiation field and the small hydrogen-like photoionization
cross-sections.

\ion{Sc}{ii} is the dominant ion of the element under solar
atmospheric conditions, with more than 99\% of the scandium atoms
being ionized under the atmospheric conditions found in the Sun. Its
lines are substantially stronger than those of \ion{Sc}{i}, although
not comparable in line strength with other metals. Only the
strongest lines form outside $\log\tau \simeq -1$. Figure \ref{fig3}
shows that all \ion{Sc}{ii} lines in the visible spectrum are formed
near to LTE conditions. There is no clear indication how to select
the proper scaling factors. In view of the minor changes due to
parameter variation, we choose typical factors to establish the
standard model atom (see Fig. \ref{fig2}). Since this choice may
affect the line formation in stars different from the Sun, we will
extend the analysis for metal-poor stars to obtain a more
significant choice of the scaling factors.

\section{Analysis of scandium lines in the solar spectrum}

In this section, we investigate the formation of \ion{Sc}{i} and
\ion{Sc}{ii} lines in the solar atmosphere and derive the scandium
abundance based on spectrum synthesis. Lines in the solar spectrum
are calculated using the plane-parallel hydrostatic MAFAGS-ODF solar
model atmosphere with $T_{\rm eff} = 5780$ K, $\log g = 4.44$,
[Fe/H] = 0.00, $\xi_t = 0.90$ km s$^{-1}$ (for a more detailed
comparison with \emph{opacity sampling} models see Grupp
\cite{gr04}). An initial scandium abundance of log
$\varepsilon_{\odot}$(Sc) = 2.99 is adopted here. It should be noted
that the atmospheric model of the Sun does \emph{not} depend on the
scandium abundance. For all elements except scandium, we assume LTE.

\subsection{Atomic line data}

For the solar abundance analysis we selected 4 \ion{Sc}{i} and 17
\ion{Sc}{ii} lines, which ideally should satisfy the following
conditions: they are relatively free from blends, and oscillator
strengths and hyperfine splitting parameters are available. However,
both conditions are not always guaranteed. In particular, HFS data
for the ground state of \ion{Sc}{ii} are missing making the analysis
of the corresponding lines more uncertain. Unfortunately, the number
of sufficiently strong lines in the visible of both ions is very
limited, and in metal-poor stars this requires concentration on the
leading lines of \ion{Sc}{ii}, since the equivalent width of
\ion{Sc}{i} 4023 in typical turnoff stars is less than 0.5 m\AA.

\begin{figure}
\centering
\resizebox{\columnwidth}{!}{\includegraphics{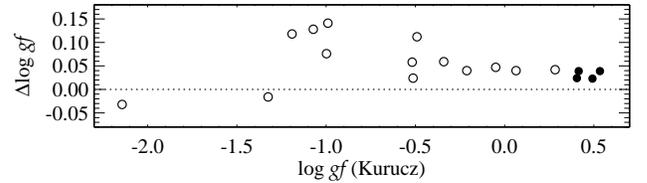}}
\caption[]{Comparison of $\log gf$-values from different sources.
Filled circles represent the \ion{Sc}{i} ion, open circles refer
to \ion{Sc}{ii}. $\Delta\log gf$ means Kurucz -- Lawler \&
Dakin.}\label{fig4}
\end{figure}
To determine the solar abundance, it is necessary to know the
accurate oscillator strengths ($f$ values) of the spectral lines.
Two sets of oscillator strengths are applied and compared in our
abundance determinations: (i) theoretical values taken from
Kurucz' database\footnote{http://kurucz.harvard.edu, see also
Kurucz \& Bell (\cite{kb95})}, and (ii) experimental data of
Lawler \& Dakin (\cite{ld89}), which were based on lifetimes
measured by Marsden et al. (\cite{ma88}) together with branching
fractions. Van der Waals damping constants log $C_6$ are computed
according to the Anstee \& O'Mara (\cite{AO91}, \cite{AO95})
interpolation tables. Input parameters needed to perform spectrum
synthesis for the selected lines are given in Table 1. The
different sets of $f$-values from Kurucz and from Lawler \& Dakin
are compared in Fig. \ref{fig4}. There is a small systematic
offset between the two data sets, with a mean $\Delta\log gf({\rm
Kur-LD}) \simeq 0.05$. The corresponding line data for our line
fits are reproduced in Table1.

In solar system matter, scandium is represented only by the
$^{45}$Sc isotope. Similar to other odd-Z elements, hyperfine
structure interactions between nuclear and electronic wave functions
split the absorption lines of Sc into multiple components. Hyperfine
structure components of line transitions are calculated as usual
from magnetic dipole splitting constants, A($J$), and electric
quadrupole splitting constants, B($J$), of the corresponding levels.
For most of the lines the HFS components fall into small intervals;
we therefore combine all components within 5 m\AA\ (\ion{Sc}{i}) or
10 m\AA\ (\ion{Sc}{ii}), which reduces many HFS patterns to two or
three coadded lines. The basic data are given in Table \ref{table2}.
Abundance differences with respect to the full HFS pattern are all
within 0.01 dex.

\begin{table}
\caption{Hyperfine structure data for
\ion{Sc}{i} and \ion{Sc}{ii} levels used in this analysis. }
\label{table2}
\begin{center}
\tabcolsep0.16cm
\begin{tabular}{lrrrl}
\noalign{\smallskip}\hline\noalign{\smallskip}
 Level & $E$ [eV] &  A$(J)$$^*$~~  &  B$(J)$$^*$~~  &  Reference \\
\noalign{\smallskip}\hline\noalign{\smallskip}
\ion{Sc}{i} & & & &  \smallskip   \\
 \Sc{a}{2}{D}{ }{5/2}  &   0.0209 &  0.00363   & --0.00124   &  Childs (1971) \\
 \Sc{a}{4}{F}{ }{7/2}  &   1.4395 &  0.0083~~  & --0.0003~~  &  Ertmer \& Hofer (1976) \\
 \Sc{a}{4}{F}{ }{9/2}  &   1.4478 &  0.0095~~  & --0.0004~~  &  Ertmer \& Hofer (1976) \\
 \Sc{z}{4}{G}{o}{9/2}  &   3.6191 &  0.0029~~  &             &  Ertmer \& Hofer (1976) \\
 \Sc{z}{4}{G}{o}{11/2} &   3.6332 &  0.0015~~  &             &  Ertmer \& Hofer (1976) \\
\noalign{\smallskip}\hline\noalign{\smallskip}
\ion{Sc}{ii} & & & &  \smallskip   \\
 \Sc{3d^2}{3}{F}{ }{3} &   0.6054 &  0.00379   & --0.00042   &  Mansour et al. (1989)  \\
 \Sc{3d^2}{3}{F}{ }{4} &   0.6184 &  0.00128   & --0.00055   &  Mansour et al. (1989)  \\
 \Sc{3d^2}{1}{D}{ }{2} &   1.3570 &  0.00498   &  0.00026    &  Mansour et al. (1989)  \\
 \Sc{3d^2}{3}{P}{ }{1} &   1.5004 & --0.00359  & --0.00041   &  Mansour et al. (1989)  \\
 \Sc{3d^2}{3}{P}{ }{2} &   1.5070 & --0.00093  &  0.00074    &  Mansour et al. (1989)  \\
 \Sc{3d^2}{1}{G}{ }{4} &   1.7682 &  0.00451   & --0.00212   &  Mansour et al. (1989)  \\
 \Sc{4p}{1}{D}{o}{2}   &   3.2337 &  0.00719   &  0.00060    &  Arnesen et al. (1982)  \\
 \Sc{4p}{3}{F}{o}{3}   &   3.4223 &  0.00685   & --0.00233   &  Young et al. (1988)    \\
 \Sc{4p}{3}{D}{o}{2}   &   3.4742 &  0.00418   &  0.00033    &  Mansour et al. (1989)  \\
 \Sc{4p}{3}{D}{o}{3}   &   3.4915 &  0.00332   &  0.00070    &  Mansour et al. (1989)  \\
 \Sc{4p}{1}{F}{o}{3}   &   4.0109 &  0.00637   & --0.00274   &  Mansour et al. (1989)  \\
 \Sc{z}{3}{P}{o}{1}    &   3.6876 &  0.00851   &  0.00033    &  Mansour et al. (1989)  \\
 \Sc{z}{3}{P}{o}{2}    &   3.6977 &  0.00354   &  0.00067    &  Mansour et al. (1989)  \\
 \Sc{4p}{3}{D}{o}{1}   &   3.4614 &  0.01016   &  0.00015    &  Mansour et al. (1989)  \\
 \Sc{4p}{3}{D}{o}{2}   &   3.4742 &  0.00418   &  0.00033    &  Mansour et al. (1989)  \\
 \Sc{4p}{3}{D}{o}{3}   &   3.4915 &  0.00332   &  0.00070    &  Mansour et al. (1989)  \\
\noalign{\smallskip}\hline\noalign{\smallskip}
\end{tabular}
\end{center}
$^{*}$: in units of cm$^{-1}$.
\end{table}

\subsection{Line profile fitting}

The observed solar flux spectrum was taken from the Kitt Peak Atlas
(Kurucz et al. \cite{KF84}). Spectrum synthesis was employed to
determine the abundance of scandium in the solar atmosphere. As in
earlier work we used the interactive spectral line-profile fitting
program SIU, which is an IDL/Fortran software package (Reetz
\cite{re91}). To match the observed spectral lines, the synthetic
spectra were convolved with a mean solar rotational velocity of 1.8
km s$^{-1}$ and a radial-tangential macroturbulence $\Xi_{\rm rt}$
which is found to vary for lines of different mean depth of
formation between 2.8 and 4.0 km s$^{-1}$.

Except for the obvious influence of strong metal or Balmer line
wings in the spectral range under consideration, we reset the local
continuum position interactively to the maximum flux in a $\pm 5
\AA$ interval around the line center. This is never a problem,
because our spectrum synthesis includes all important lines and thus
allows confirmation of the local maximum flux estimate. Any
uncertainty in this process, even in the yellow wavelength range
between 5700 and 5850 \AA~ where the solar atlas displays a
continuum that is systematically high by $\sim 2\%$, is smaller than
the general profile fitting error. Our estimate of its influence on
the abundances is $\sim 0.01$ dex.

Line profiles are computed under both LTE and NLTE assumptions:
fitted to the observed profiles by means of scandium abundance
variations. Column 10 of Table \ref{table1} reproduces the
logarithmic abundances $\log \varepsilon$ of the fits including a
number of weak blends on either side of the profiles. They are
based on Kurucz' log $gf$ values and NLTE level populations. The
logarithmic corrections due to weak blends are listed in the
$\Delta b$ column. Thus, fitting the lines \emph{without} blends
would have resulted in a \emph{higher} logarithmic abundance $\log
\varepsilon - \Delta b$. The difference required to fit LTE and
NLTE profiles is referred to as the {\it NLTE correction} ($\Delta
X = \log \varepsilon^{\rm NLTE} - \log\varepsilon^{\rm LTE}$); it
is given in column 13 of Table \ref{table1} for each line.
Equivalent widths from NLTE profile fits (see Fig. \ref{fig5}) are
given in column 6 of Table \ref{table1}. The last three columns
give the number of components, the maximum wavelength separation
(in m\AA) of the hyperfine structure (HFS) lines and an asterisk,
if HFS data are missing for one of the levels(cf. Table
\ref{table2}).

\begin{figure*}
\resizebox{\textwidth}{!}{\includegraphics{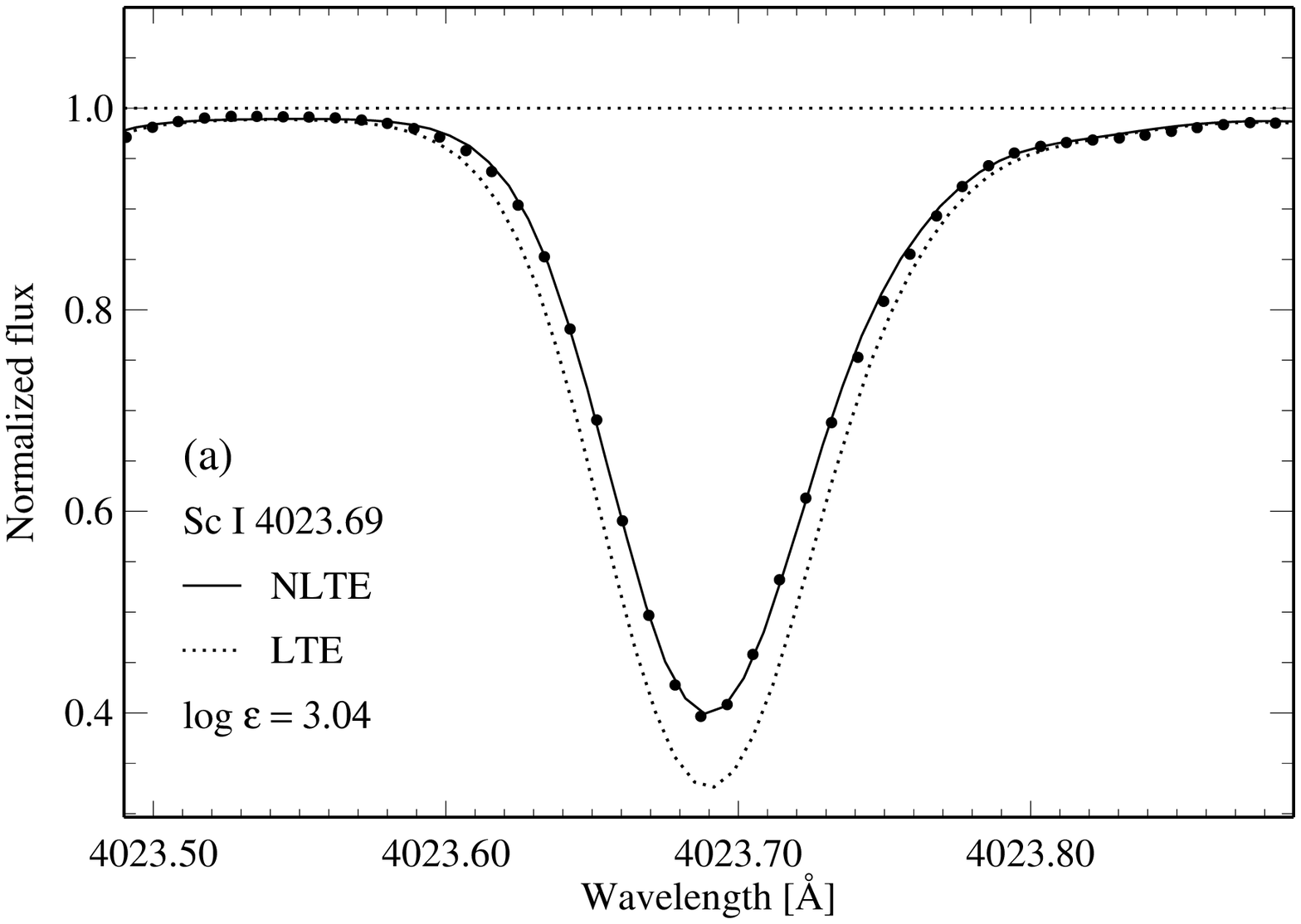}\includegraphics{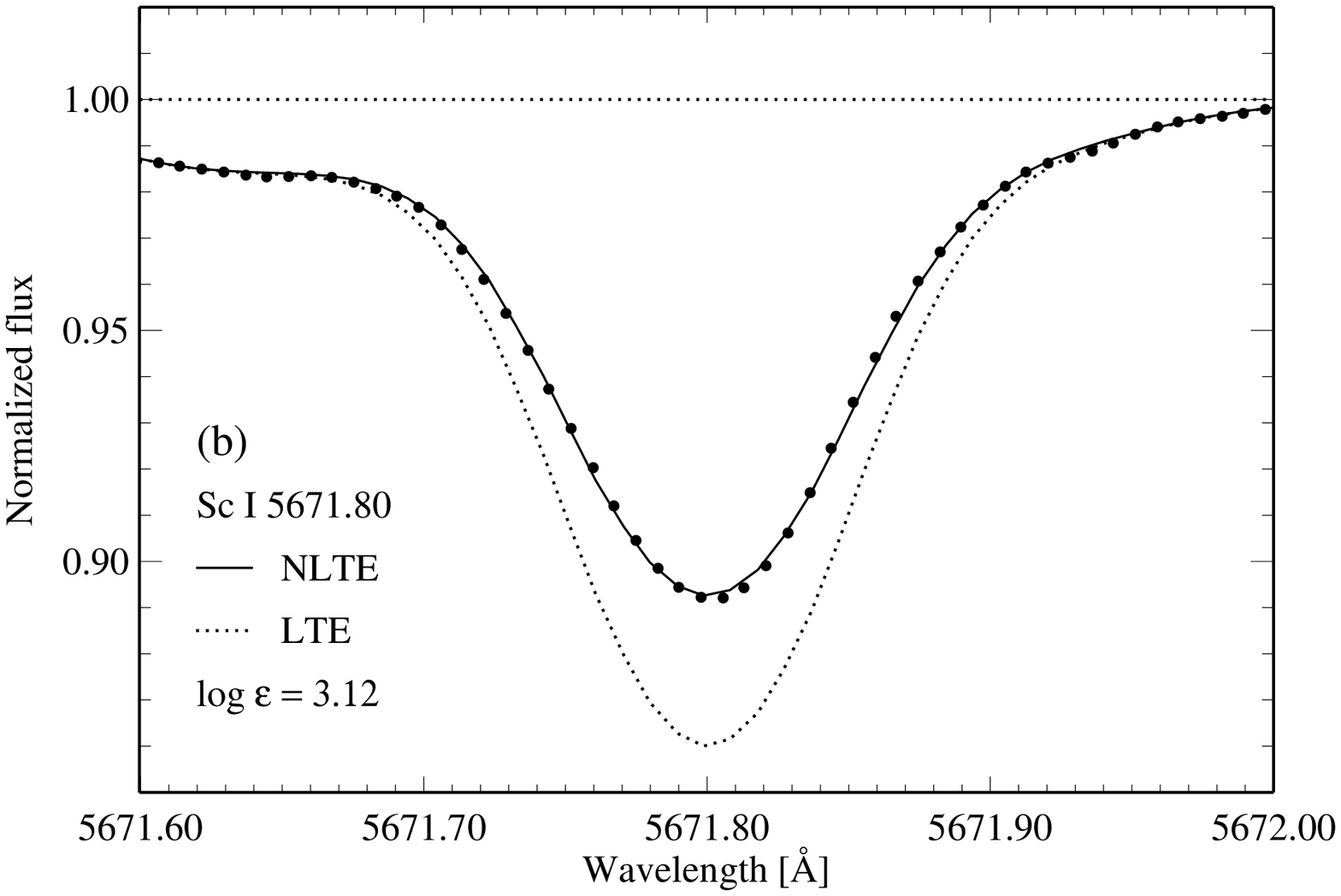}}\\
\vspace{-0.cm}
\resizebox{\textwidth}{!}{\includegraphics{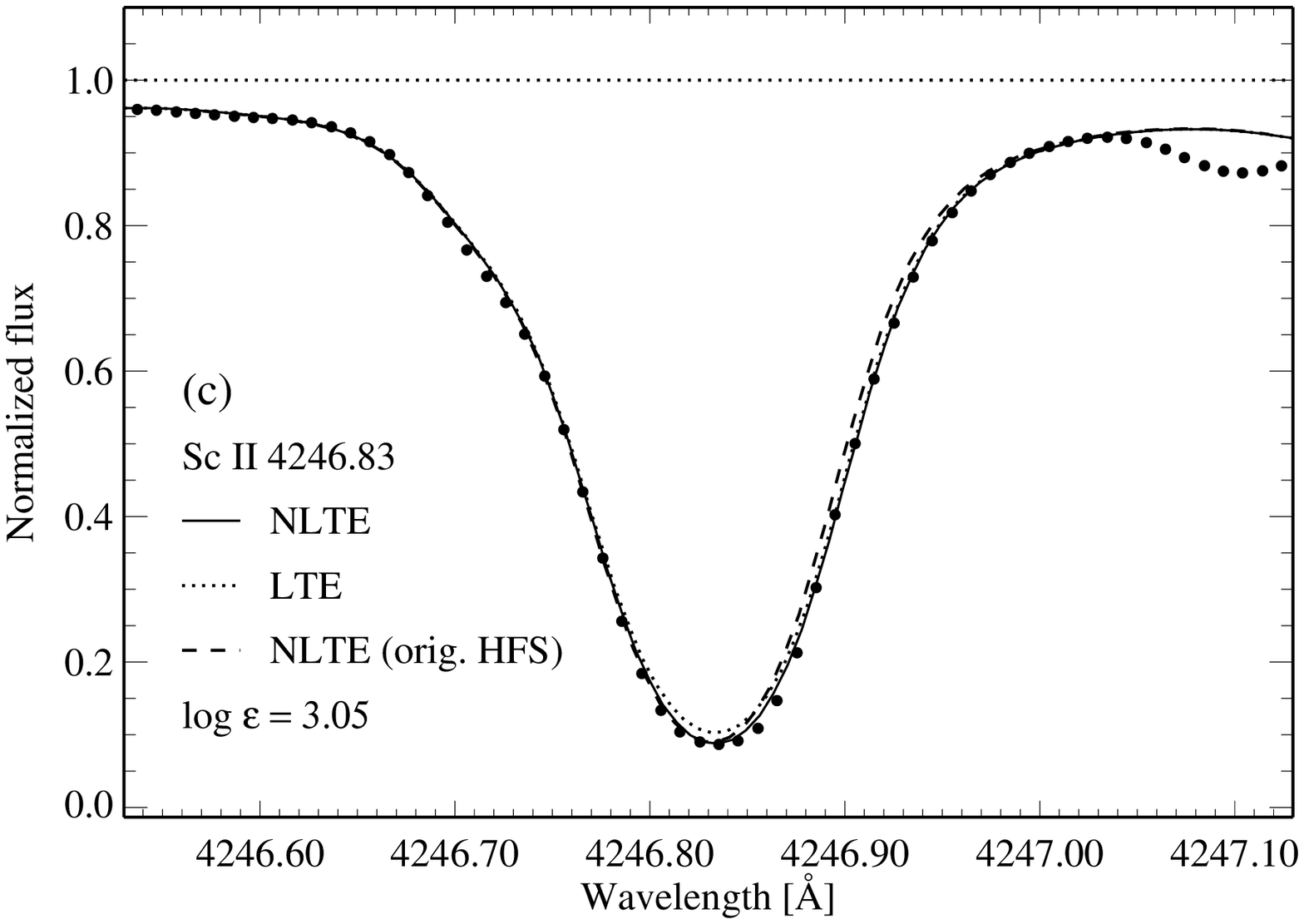}\includegraphics{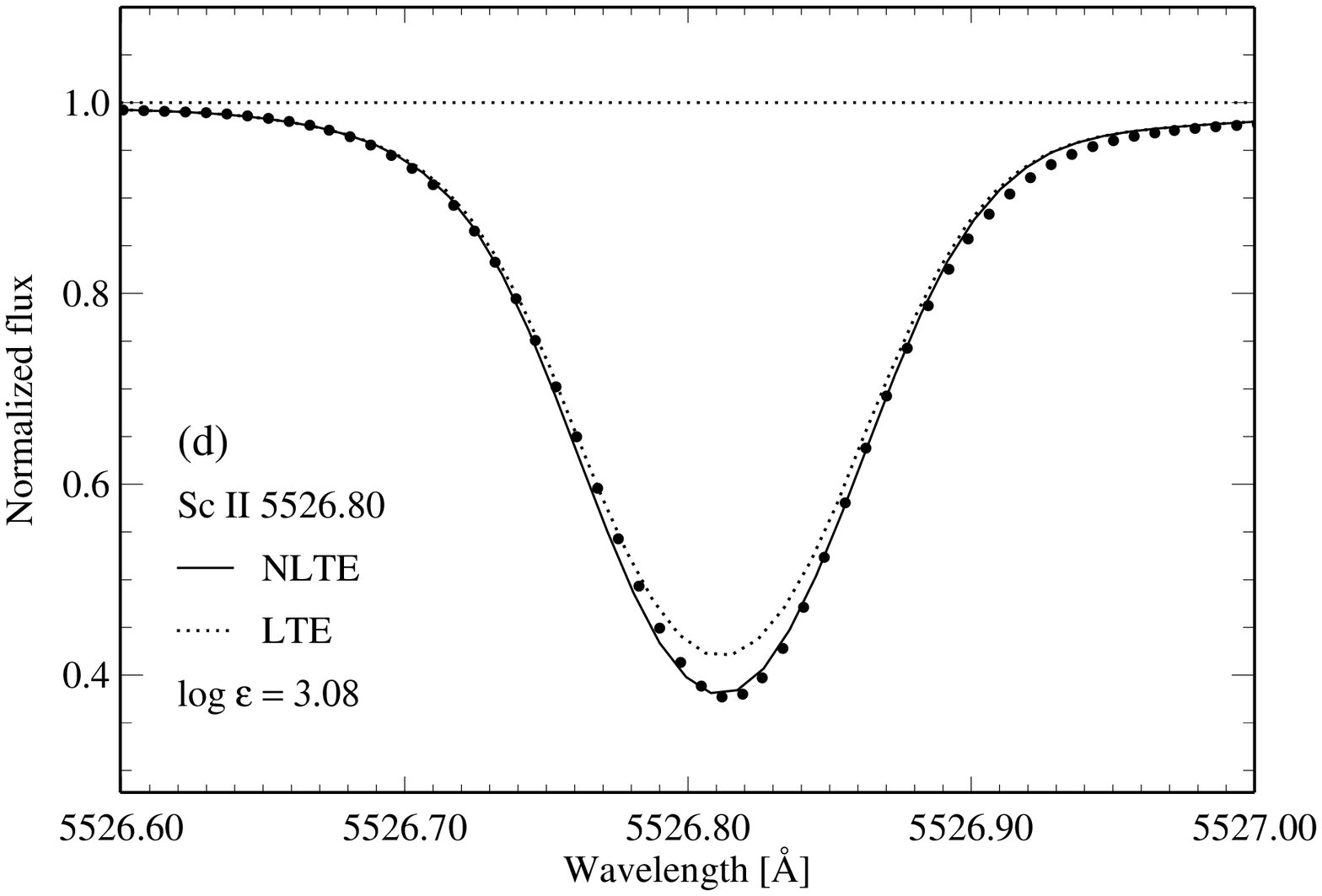}}\\
\vspace{-0.cm}
\resizebox{\textwidth}{!}{\includegraphics{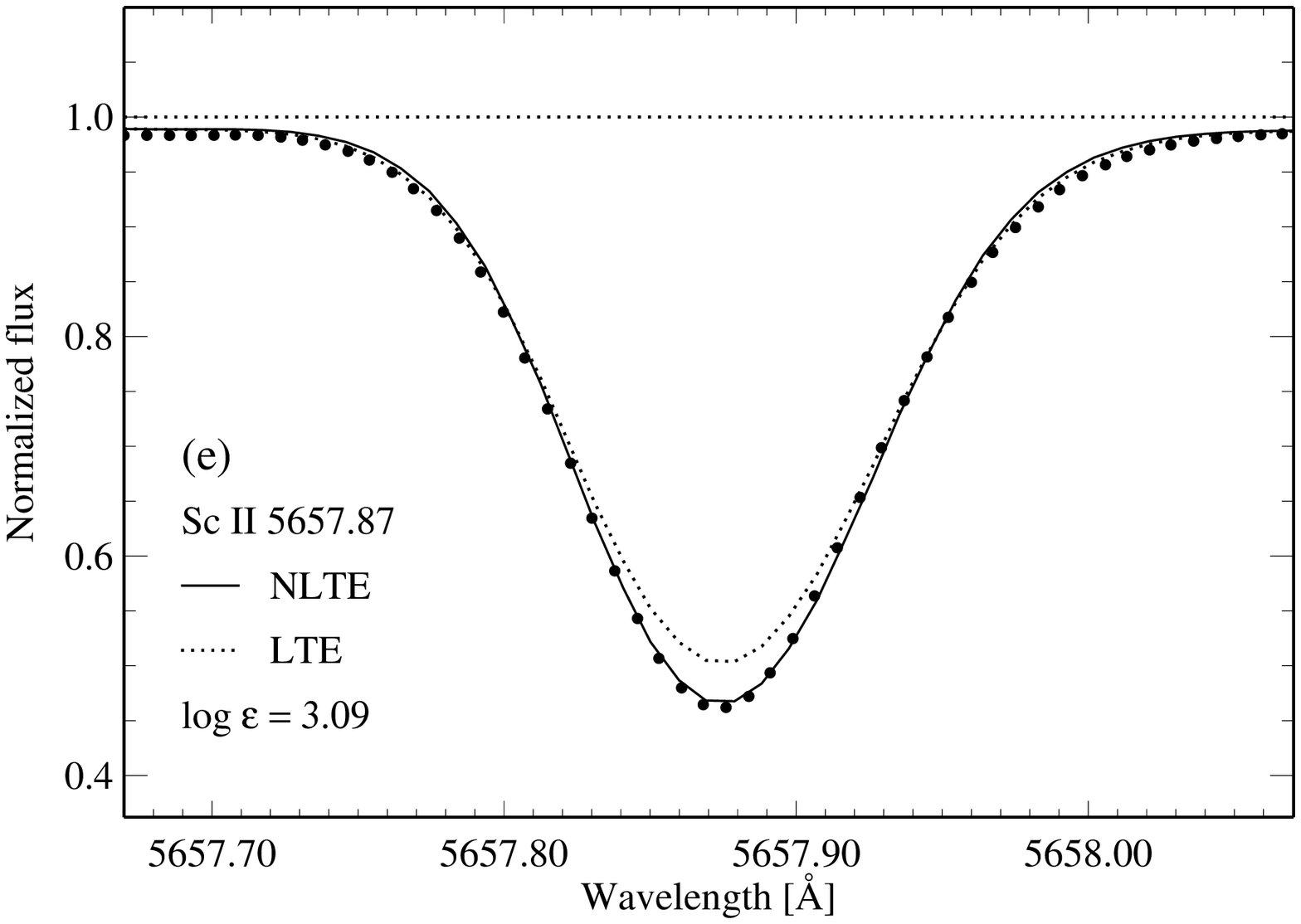}\includegraphics{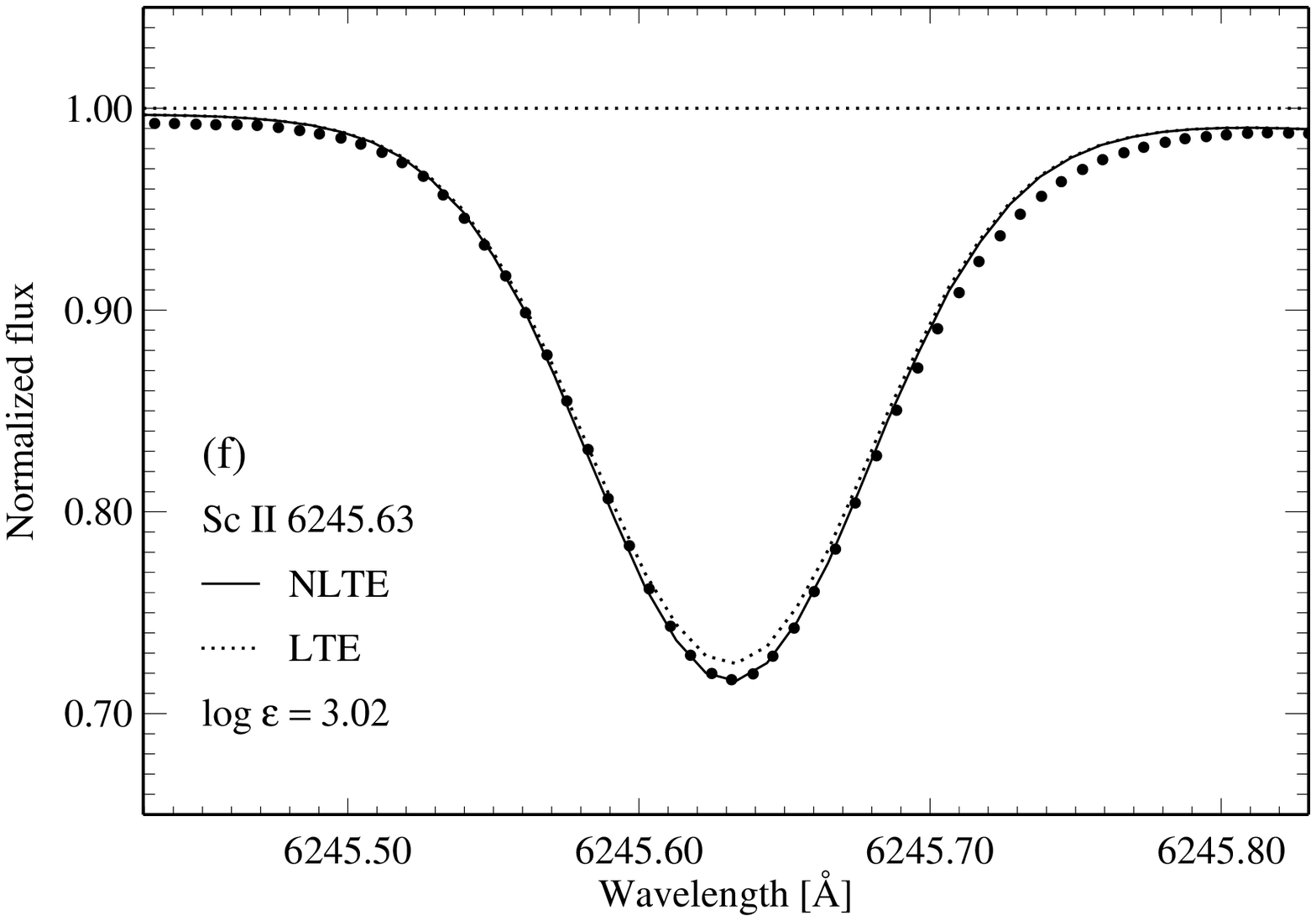}}
\vspace{-0.5cm} \caption[]{Representative solar line profiles for
\ion{Sc}{i} and \ion{Sc}{ii}. Always shown are the NLTE and LTE
profiles for the given abundance, where the NLTE profiles refer to
the best fit. $\lambda\ 4246.83$ additionally shows the NLTE line
with the \emph{originally calculated} HFS components (second
component at 4246.839 instead of 4246.844 \AA, see
text).}\label{fig5}
\end{figure*}

Some of the synthetic profiles for selected lines, together with the
observed solar spectrum, are presented in Fig. \ref{fig5}. Whenever
available, we have included known blends. For comparison we show
profiles under LTE and NLTE conditions including HFS. Generally, all
neutral lines are much fainter in NLTE due to significant
underpopulation, whereas lines of ionized scandium are slightly
stronger under NLTE conditions.

A marginal \emph{line core asymmetry} with the observed line
bisectors shifted 1 or 2 m\AA\ to the red seems to exist. This is
known from the solar spectrum synthesis of other metals, such as Si,
Ca, or Mn. It is probably the result of hydrodynamic streaming
patterns that cannot be represented by our simple
micro-/macroturbulence scheme. A more disturbing defect is the
near-continuum flux deficiency of the red line wing, best seen in
the 5657 and 6245 \AA\ lines. This feature is also found in other
metals. Sometimes addressed as the result of a weak line haze or
blends, the systematic blue-red asymmetry is more likely to result
from hydrodynamic flows, too. We emphasize, however, that this
wavelength region displays a particularly disturbed solar continuum
flux. } Table \ref{table1} already gives a hint that some of the
lines synthesized here suffer from missing HFS data. Unfortunately,
the strongest lines of both ions are affected, requiring more
detailed comments.

\ion{Sc}{i} 4023.69 \AA\ has been calculated with only the HFS split
of the \Sc{a}{2}{D}{ }{5/2} level. That this is possibly a fair but
not perfect representation of the true hyperfine structure width of
the line is documented in Fig. \ref{fig5}a, where the synthesized
line halfwidth fits that of the solar spectrum. However, the total
separation of the HFS components is small (see Table \ref{table1}).
Another distortion is caused by a number of faint line blends on the
red core and wing of this line. The known components are two highly
excited faint \ion{Mn}{i} lines at 4023.72 and 4023.84 \AA, for
which no HFS data are available. There is also a faint \ion{Cr}{i}
line at 4023.74 \AA. These blends have been considered in Fig.
\ref{fig5}a, but even the two lines within 50 m\AA\ of the
\ion{Sc}{i} line center have no influence on the fit of the core.
Introducing these blend components and fitting the full core profile
results in a Sc line abundance change below $0.01$ dex. The
remaining red wing depression centered on 4023.84 \AA\ may be the
result of the unknown \ion{Mn}{i} HFS.

\ion{Sc}{i} 5081.55, 5671.80, and 5686.83  \AA\ are substantially
fainter neutral Sc lines, with solar equivalent widths of only 12,
14, and 9 m\AA, respectively. That makes the line abundances be
sensitive to the continuum position, which is particularly uncertain
between 5600 and 5700 \AA. Only one of the lines is presented in
Fig. \ref{fig5}b. Although most of the Sc lines were chosen to be as
free of blends as possible, the wings of the 5671.80 \AA\ line are
covered by quite a few weak lines of \ion{Mn}{i}, \ion{Fe}{i}, and
\ion{Ti}{i}, all of which are in the range of equivalent widths
between 0.5 and 1.5 m\AA. These lines have central depths between 1
and 2\% of the continuum flux, and their (gaussian) wings give some
combined contribution to the Sc line core. Thus the full influence
of the weak blends on the Sc abundance of this line is $-0.04$ dex.
For the other two lines, the blend corrections are $-0.04$ and
$-0.07$, respectively.

\ion{Sc}{ii} 4246.83 \AA\ is one of the strongest lines of
\ion{Sc}{ii} in the visible. It contains two weak blends on the blue
line wing. Due to missing HFS data for the lower level,
\Sc{a}{1}{D}{ }{2}, the resulting profiles are not as reliable as
those obtained for the excited levels. This is evident from Fig.
\ref{fig5}c, where the line width of the profile with the originally
calculated HFS components does not fit the observed solar profile
unless the HFS separation is increased by 5 m\AA\ moving the second
component from 4246.839 \AA\ to 4246.844 \AA. This is a purely
empirical correction; however, it is the only way to fit the solar
line profile without postulating an unknown blend.

\ion{Sc}{ii} 5526.81 \AA\ and 5657.87 \AA\ (Figs. \ref{fig5}d and e)
represent the lines on the flat part of the curve-of-growth; i.e.,
they strongly depend on the microturbulence parameter. Their
hyperfine structure is known and seems to fit the solar spectrum for
both lines. There are two blends on the red wing of the 5526 \AA\
line, but not near enough to the core of the Sc line to affect the
abundance. A few weak neutral Cr, Fe, and V lines on both wings of
the 5657 \AA\ line have no influence on the abundance either.

\ion{Sc}{ii} 6245.63 \AA\ belongs to the group of well-separated,
unblended weak lines. Again, the hyperfine structure fits the
observed spectra as shown in Fig. \ref{fig5}f. This line lies on the
extended wing of \ion{Fe}{i} (816) 6246.334 \AA, but that does not
change the line abundance.

Altogether, the remaining influence of blends is small, with
$\overline{\Delta b} = -0.021$, but it is systematic in that it
always \emph{reduces} the mean Sc abundance. It is even more
important because it tends to reduce the peaks of the abundance
distribution.

\subsection{The solar scandium abundance}

\begin{figure}
\hbox{\resizebox{\hsize}{!}{\includegraphics{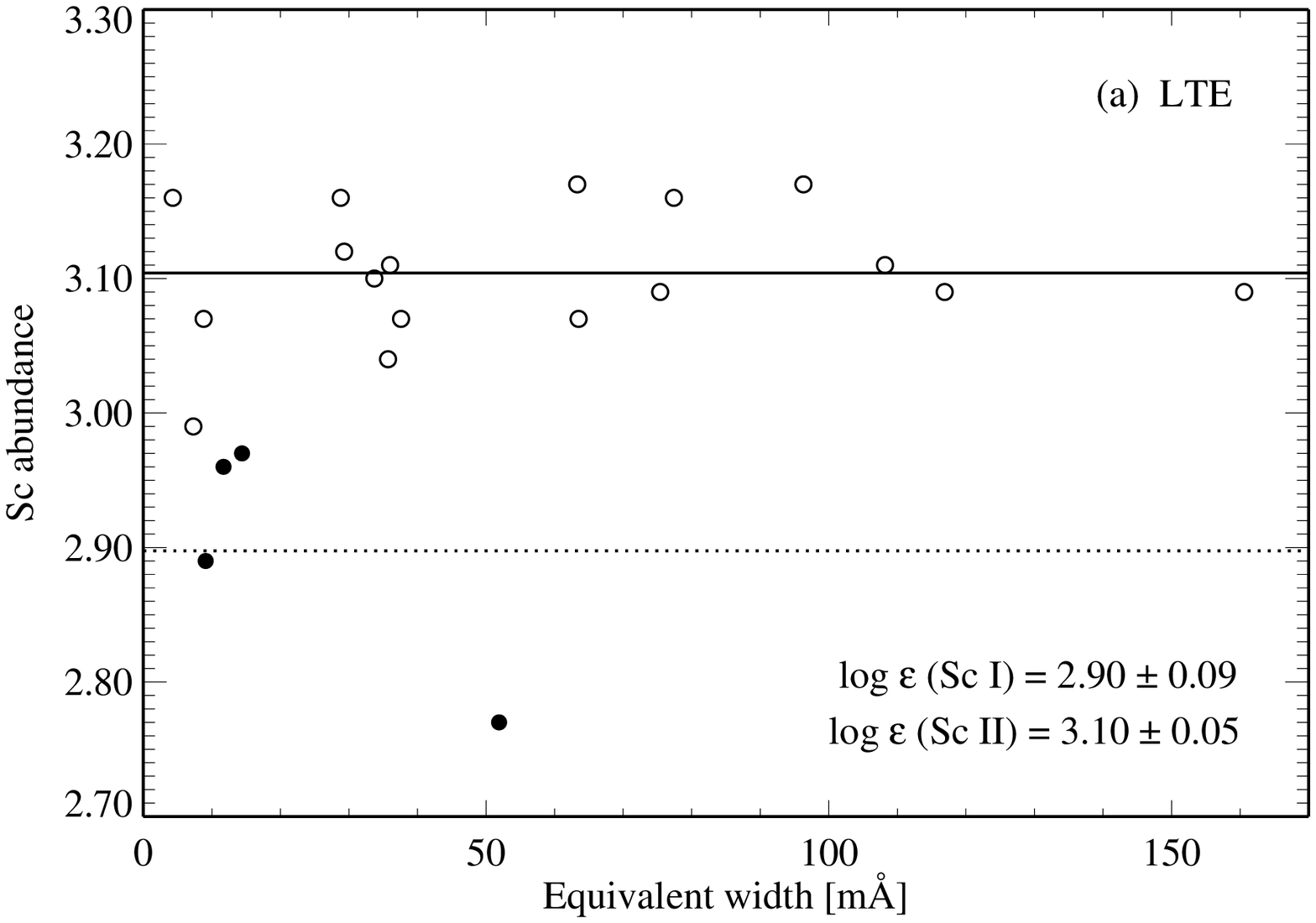}}}
\vspace{0mm}
\hbox{\resizebox{\hsize}{!}{\includegraphics{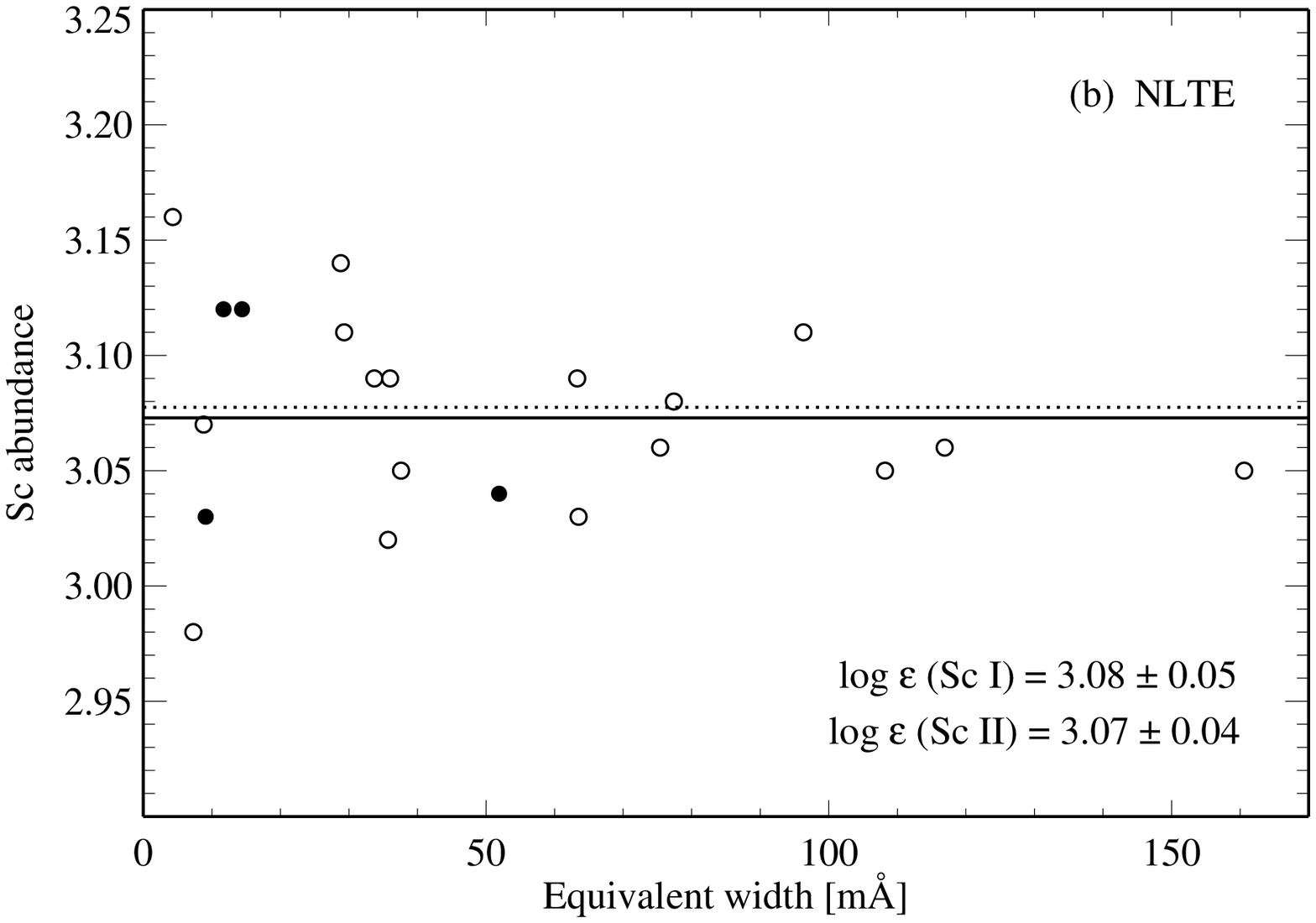}}}
\caption[]{Solar Sc abundances based on Kurucz $gf$ values for
lines of \ion{Sc}{i} (filled circles) and \ion{Sc}{ii} (open
circles) plotted as function of the line equivalent widths. The
mean abundance from \ion{Sc}{i} lines is represented by the dotted
line, that of \ion{Sc}{ii} by a continuous line. Top: LTE, bottom:
NLTE.} \label{fig6}
\end{figure}

Our method of spectrum synthesis yields the product of the
oscillator strength for a given transition and the abundance of the
element, log $gf \varepsilon_{\odot}$. The results are reproduced in
Table 1 assuming NLTE conditions. Using the values obtained for log
$gf \varepsilon_{\odot}$ and the $\log gf$ values from different
data sets, we computed Sc abundances for the individual lines.
Figure 6 shows LTE and NLTE abundance results based on the Kurucz
calculated oscillator strengths for all lines as a function of their
equivalent widths.

Under the LTE assumption, absolute solar abundances determined from
\ion{Sc}{i} lines are significantly lower than the values obtained
for the \ion{Sc}{ii} lines. The \emph{mean} LTE abundances for 4
\ion{Sc}{i} and 17 \ion{Sc}{ii} lines are $2.90 \pm 0.09$ dex and
$3.10 \pm 0.05$ dex, respectively. This discrepancy of the
ionization equilibrium is resolved in NLTE calculations. Under NLTE
assumption, \ion{Sc}{i} and \ion{Sc}{ii} lines give very consistent
abundance results, i.e. $3.08 \pm 0.05$ dex and $3.07 \pm 0.04$ dex,
respectively. Using Kurucz' $gf$ values, the mean abundance for all
21 Sc lines under NLTE is $3.07 \pm 0.04$ dex.

Using instead the laboratory $\log gf$ values of Lawler \& Dakin, the mean
abundance of 17 Sc lines under NLTE is $3.13 \pm 0.05$ dex. The ionization
equilibrium differs by 0.03 dex. While the results for both sets of $gf$ values
are a brilliant justification of the NLTE assumption itself, it seems that the
experimental $gf$ values provided by Lawler \& Dakin (\cite{ld89}) should be
slightly more reliable than those calculated by Kurucz. However, the standard
deviation of the experimental data is slightly higher.

\section{Discussion}
The most interesting result of this analysis turns out to be that
scandium is the first element found to show strong solar NLTE
\emph{abundance} effects in one of its ions. This appears to be the
consequence of missing strong lines in \ion{Sc}{i}. All the lines of
this ion are weak because of two contributions: (a) the solar
scandium abundance is ~3 dex lower than that of other metals, and
(b) due to its low ionization energy of only 6.6 eV, \ion{Sc}{i} is
an extreme \emph{minority ion}, that is 99.8\% ionized in its solar
region of line formation. Therefore it does not help that the $gf$
values are relatively normal, and the statistical equilibrium of
\ion{Sc}{i} is far from thermal even in layers of continuum
formation. The departure coefficients calculated for different
values of photoionization and collision parameters (see Fig.
\ref{fig3}) reveal the insensitivity of our results to the choice of
details of the model atom. This may no longer be the case in more
metal-poor stars.

Although still far from complete, the experimental \emph{hyperfine
structure} data are compatible with the observed solar line
profiles, with the exception of a few lines, for which either lower
or upper level HFS data are missing. We have not attempted to
determine abundances without HFS, because the profile fits were far
from realistic in most cases (in particular for \ion{Sc}{ii}). The
general trend of such an analysis would be a relative
\emph{increase} in the abundances as a response to a decrease in
line broadening, which affects mostly the few lines on the flat part
of the curve-of-growth.

Depending on whether calculated $gf$ values of Kurucz \& Bell (\cite{kb95}) or
experimental $gf$ values of Lawler \& Dakin (\cite{ld89}) are preferred, the
solar scandium abundance is
$$ \log \varepsilon_{{\rm Sc},\odot} = \left\{ \begin{array}{l}
                                      3.07 \pm 0.04 \qquad {\rm (Kurucz\ \&\ Bell)}\\
                                      3.13 \pm 0.05 \qquad {\rm (Lawler\ \&\ Dakin )~~.}
                                      \end{array} \right. $$

For the experimental $gf$ values, the difference with respect to
the \emph{meteoritic} scandium abundance turns out to be as high
as before. Therefore the emphasis lies on our current analysis
being much more restrictive than recent results. If the higher
photospheric abundance were caused by a systematic error in our
analysis, this should be found in Fig. \ref{fig6}b. Assuming say
that all lines were affected by remaining unknown blends, the
resulting abundances should show a trend with line strength (weak
line abundances are more affected than strong lines). Since such a
trend is not found, we may rule out the importance of unknown
blends. A similar argument holds for the role of the HFS, however,
now producing an inverse trend, where weak lines should not depend
on the HFS.

A completely different systematic abundance error could result from
the solar model atmosphere. In the above investigation we started
the statistical equilibrium and the synthetic spectrum calculations
with our standard ODF model atmosphere\footnote{This should not lead
to confusion, because the background line opacities in the
statistical equilibrium calculations are \emph{always} sampled.},
which is very much the same as that of Kurucz (\cite{ku92}), and it
makes use of his opacity distribution functions, corrected for metal
abundance. Such ODF models tend to have slightly lower temperatures
near optical depth $\tau \simeq 1$. Opacity sampling (OS) models are
different. Their higher temperatures are essentially the reason for
the insufficient fit the solar Balmer lines (see Grupp \cite{gr04}).
Since the temperature difference for the solar OS model of Grupp is
only around 40 K or less, it changes the Sc line formation by a
negligible amount. The abundance entries in Table \ref{table1}
document that there is a mean abundance difference in the sense
ODF$-$OS of only --0.016 dex.

Comparison with the results obtained for the ODF model atmosphere
shows that the solar scandium abundance for the laboratory $gf$
values of Lawler \& Dakin, $\log \varepsilon_{{\rm Sc},\odot} = 3.13
\pm 0.05$, is off the meteoritic value of $3.04$ by 0.09 dex. A
large fraction of the remaining scatter of the single line
abundances seen in Fig. \ref{fig6}b is probably caused by the
uncertain $gf$ values, both for calculated and experimental data.
While previous analyses could not confirm a correspondence between
photospheric and meteoritic Sc abundances due to a relatively large
line-by-line scatter, our results reduce the problem to a simple
discrepancy. Kurucz' $f$-values lead to a solar Sc abundance well in
agreement with the meteoritic value, whereas the experimental data
of Lawler \& Dakin deviate from that reference by nearly $2\sigma$.
Currently, there is no hint as to why photospheric scandium should
differ in abundance from that found in chondrites.

\begin{acknowledgements}
This project was supported by the Deutsche Forschungsgemeinschaft
(DFG) under grants GE490/33-1 and 446 CHV 112/1,2/06, and by the
National Natural Science Foundation of China under grants No.
10778612, 10433010, and 10521001, and by the National Key Basic
Research Program (NKBRP) No. 2007CB815403. HWZ and GZ thank the
Institute of Astronomy and Astrophysics of Munich University for
warm hospitality during a productive stay in 2006 and 2007.
\end{acknowledgements}


\begin{thebibliography}{}

\bibitem[1973]{al73} Allen, C.W. 1973, Astrophysical Quantities, 3rd Ed., Athlone Press, London
\bibitem[1989]{ag89} Anders, E., \& Grevesse, N. 1989, \gca, 53, 197
\bibitem[1991]{AO91} Anstee, S.D., \& O'Mara, B.J. 1991, MNRAS, 253, 549
\bibitem[1995]{AO95} Anstee, S.D., \& O'Mara, B.J. 1995, MNRAS, 276, 859
\bibitem[1982]{AH82} Arnesen, A., Hallin, R., Nordling, C., et al. 1982, \aap, 106, 327
\bibitem[2005]{as05} Asplund, M. 2005, \araa, 43, 481
\bibitem[2000]{bpo00} Barklem, P.S., Piskunov, N., O'Mara, B.J. 2000, A\&AS, 142, 467
\bibitem[1999]{be99} Belyaev, A.K., Grosser, J., Hahne, J., et al. 1999, Phys.Rev.A, 60, 2151
\bibitem[2003]{bb03} Belyaev, A.K., \& Barklem, P.S. 2003, Phys.Rev.A, 68, 2703
\bibitem[1952]{BK52} Brix F., \& Kopfermann H. 1952, Landolt-B\"ornstein, Zahlenwerte und Funktionen I/5, Springer, Berlin
\bibitem[1985]{BG85} Butler K., \& Giddings J. 1985, Newsletter on the analysis os astronomical spectra No. 9, University of London
\bibitem[1968]{C71} Childs, W.J. 1971, Phys.Rev.A, 4, 1767
\bibitem[1968]{dr68} Drawin, H.W. 1968, Z.Physik, 211, 404
\bibitem[1969]{dr69} Drawin, H.W. 1969, Z.Physik, 225, 483
\bibitem[1976]{EH76} Ertmer, W., \& Hofer, B. 1976, Z.Physik, 276, 9
\bibitem[1997]{fu97} Fuhrmann, K., Pfeiffer, M., Frank, C., et al. 1997, \aap, 323, 909
\bibitem[2001]{GBMRS01} Gehren, T., Butler, K., Mashonkina, L., et al. 2001, \aap, 366, 981
\bibitem[1984]{g84} Grevesse, N. 1984, Phys.Scr., 8, 49
\bibitem[1993]{gn93} Grevesse, N., \& Noels, A. 1993, in Origin and Evolution of the Elements, ed. N. Prantzos, E. Vangioni-Flam, \& M. Casse (Cambridge: Cambridge Univ. Press), 15
\bibitem[2007]{gas07} Grevesse, N., Asplund, M., Sauval, A.J. 2007, Space Sci. Rev., 130, 105
\bibitem[1991]{gs91} Gratton, R.G., \& Sneden, C. 1991, \aap, 241, 501
\bibitem[2004]{gr04} Grupp, F. 2004, \aap, 420, 289
\bibitem[1974]{HM74} Holweger, H., \& M\"uller, E.A. 1974, Sol. Phys., 39, 19
\bibitem[1984]{KF84} Kurucz, R.L., Furenlid, I., Brault, J., et al. 1984, Solar Flux Atlas from 296 to 1300nm, Kitt Peak National Solar Observatory
\bibitem[1995]{kb95} Kurucz, R.L., \& Bell, B. 1995, Atomic Line Data, Kurucz CD-ROM No. 23., Cambridge, Mass.: Smithsonian Astrophysical Observatory
\bibitem[1992]{ku92} Kurucz R.L. 1992, Rev. Mex. Astron. Astrof., 23, 45
\bibitem[1989]{ld89} Lawler, J.E., \& Dakin, J.T. 1989, J. Opt. Soc. Am. B, 6, 1457
\bibitem[1989]{MD89} Mansour, N.B., Dineen, T., Young, L., et al. 1989, Phys.Rev.A, 39, 5762
\bibitem[1988]{ma88} Marsden, G.C., Den Hartog, E.A., Lawler, J.E., et al. 1988, J. Opt. Soc. Am. B, 5, 606
\bibitem[1978]{mi78} Mihalas, D. 1978, Stellar Atmospheres, 2nd edition (San Francisco, W.H. Freeman \& Co.), p. 99
\bibitem[1993]{neu93} Neuforge, C. 1993, in Origin and Evolution of the Elements, ed. N. Prantzos, E. Vangioni-Flam, \& M. Casse, Cambridge: Cambridge Univ. Press, 63
\bibitem[2000]{ncsz00} Nissen, P.E., Chen, Y.Q., Schuster, W.J., et al. 2000, \aap, 353, 722
\bibitem[2000]{pm00} Prochaska, J.X., \& McWilliam, A. 2000, \apj, 537, L57
\bibitem[1990]{re91} Reetz, J.K. 1991, Diploma Thesis, Universit\"at M\"unchen
\bibitem[1962]{se62} Seaton, M.J. 1962, in Atomic and Molecular Processes, Acad. Press, New York
\bibitem[1994]{se94} Seaton, M.J., Mihalas, D., Pradhan, A.K. 1994, \mnras, 266, 805
\bibitem[1984]{sh84} Steenbock, W., \& Holweger, H. 1984, \aap, 130, 319
\bibitem[1962]{re62} van Regemorter H. 1962, ApJ, 136, 906
\bibitem[1988]{YC88} Young L., Childs W.J. Dineen C. et al. 1988, Phys.Rev.A, 37, 4213
\bibitem[1989]{ya89} Youssef, N.H., \& Amer, M.A. 1989, \aap, 220, 281
\bibitem[2006]{zg06} Zhang, H.W., Butler, K., Gehren, T., et al. 2006, \aap, 453, 723
\bibitem[1990]{zm90} Zhao, G., \& Magain, P. 1990, \aap, 238, 242




\end{thebibliography}
\end{document}